\documentclass[aps,prl,twocolumn,english,balance,superscriptaddress,floats,showpacs,letter, prb]{revtex4-2}
\usepackage[latin9]{inputenc}
\usepackage{physics}
\usepackage{amsmath}
\usepackage{amssymb}
\usepackage{graphicx}
\usepackage{esint}
\usepackage{subfigure}
\usepackage{multirow}
\usepackage{mathtools}
\usepackage{xcolor}
\usepackage{gensymb}
 
\makeatletter

\newcommand{\beq}{\begin{equation}}
\newcommand{\eeq}{\end{equation}}
\newcommand{\bea}{\begin{eqnarray}}
\newcommand{\eea}{\end{eqnarray}}
\newcommand{\bwt}{\begin{widetext}}
\newcommand{\ewt}{\end{widetext}}
\@ifundefined{textcolor}{}
{%
 \definecolor{BLACK}{gray}{0}
 \definecolor{WHITE}{gray}{1}
 \definecolor{RED}{rgb}{1,0,0}
 \definecolor{GREEN}{rgb}{0,1,0}
 \definecolor{BLUE}{rgb}{0,0,1}
 \definecolor{CYAN}{cmyk}{1,0,0,0}
 \definecolor{MAGENTA}{cmyk}{0,1,0,0}
 \definecolor{YELLOW}{cmyk}{0,0,1,0}
}

\newcommand{\eps}{\epsilon}
\newcommand{\bA}{\mathbf{A}}
\newcommand{\bB}{\mathbf{B}}
\newcommand{\bk}{\mathbf{k}}
\newcommand{\bq}{\mathbf{q}}
\newcommand{\bp}{\mathbf{p}}
\newcommand{\bg}{\mathbf{g}}
\newcommand{\br}{\mathbf{r}}
\newcommand{\by}{\mathbf{y}}
\newcommand{\bL}{\mathbf{L}}
\newcommand{\bK}{\mathbf{K}}

\usepackage{hyperref}
\hypersetup{colorlinks,
linkcolor={blue},
citecolor={blue},
urlcolor={purple}}

\makeatother

\begin{document}
	
\title{Narrow bands in magnetic field and strong-coupling Hofstadter spectra}

\author{Xiaoyu Wang}
\affiliation{National High Magnetic Field Laboratory, Tallahassee, Florida, 32310, USA}

\author{Oskar Vafek}
\email{vafek@magnet.fsu.edu}
\affiliation{National High Magnetic Field Laboratory, Tallahassee, Florida, 32310, USA}
\affiliation{Department of Physics, Florida State University, Tallahassee, Florida 32306, USA}

\begin{abstract}
We develop a new, efficient, and general method to determine the Hofstadter spectrum of isolated narrow bands. The method works for topological as well as for trivial narrow bands by projecting the zero $\mathbf{B}$-field hybrid Wannier states -- which are localized in one direction and Bloch extended in another direction -- onto a representation of the magnetic translation group in the Landau gauge. We then apply this method to find, for the first time, the Hofstadter spectrum for the exact single particle charged excitations in the strong coupling limit of the magic angle twisted bilayer graphene at the charge neutrality point and at $|\nu|=2$ down to low magnetic fields when the flux through the moir\'e unit cell is only $\sim 1/25$ of the electronic flux quantum i.e. $\sim 1$T at the first magic angle. The resulting spectra provide a means to investigate Landau quantization of the quasiparticles even if their dispersion is interaction induced.
\end{abstract}

\maketitle


The rise of moir\'e materials \cite{Cao2018a,Cao2018b,Kerelsky2019,Lu2019,Jiang2019,Yankowitz2019,Choi2019,Sharpe2019,Xie2019,Zondiner2020,Wong2020,Serlin2020,Stepanov2020,Liu2021,Yacoby2021,Wu2021} has brought into focus the challenge to understand the physics of correlated narrow bands subject to quantizing magnetic field $\bB$ \cite{Bistritzer2011b,Moon2014,Hejazi2019,Yahui2019,Biao2020LL,Jonah2020,Jonah2021}. Such narrow bands can be topologically non-trivial even at $\bB=0$, as is the case for the magic angle twisted bilayer graphene (MATBG)\cite{Po2018,Junyeong2019,Zhida2019}. Moreover, for a moir\'e period $\sim 13$nm, as in MATBG, the magnetic flux through the unit cell, $\phi$, can readily become comparable or even exceed the flux quantum $\phi_0=hc/e$ using existing high field magnets, so that the interplay of strong correlation and Hofstadter physics can be realized in a laboratory \cite{dean_hofstadters_2013,Saito2021,Yacoby2021,Finney2021}.

The traditional way to determine the non-interacting Hofstadter spectrum in the MATBG is to minimally couple the magnetic vector potential $\bA$ to the continuum Bistritzer-MacDonald (BM) Hamiltonian~\cite{Bistritzer2011} and then to expand it in the Landau level (LL) basis \cite{Bistritzer2011b,Moon2014,Hejazi2019,Yahui2019}. Although this provides a reliable method, it requires a large upper cutoff on the LL index \cite{Hejazi2019} in order to converge, particularly at low $\bB$, or close to simple rational values of $\phi/\phi_0=p/q$ where the LL basis method becomes prohibitively computationally expensive. This is because many Landau quantized remote bands are effectively kept together with the Landau quantized narrow bands of interest. Equivalently, at low $\bB$, the real space shape of the narrow band wavefunctions --with peaks in the local density of states at the moir\'e triangular lattice sites-- is mainly determined by the interlayer tunneling ($w_{0,1}$) induced periodic potential and a superposition of a large number of LLs is needed in order to recover such real space structure. If one is then interested in interaction induced phenomena within the resulting narrow bands a more efficient method is desirable.

\begin{figure}
    \centering
    \includegraphics[width=\linewidth]{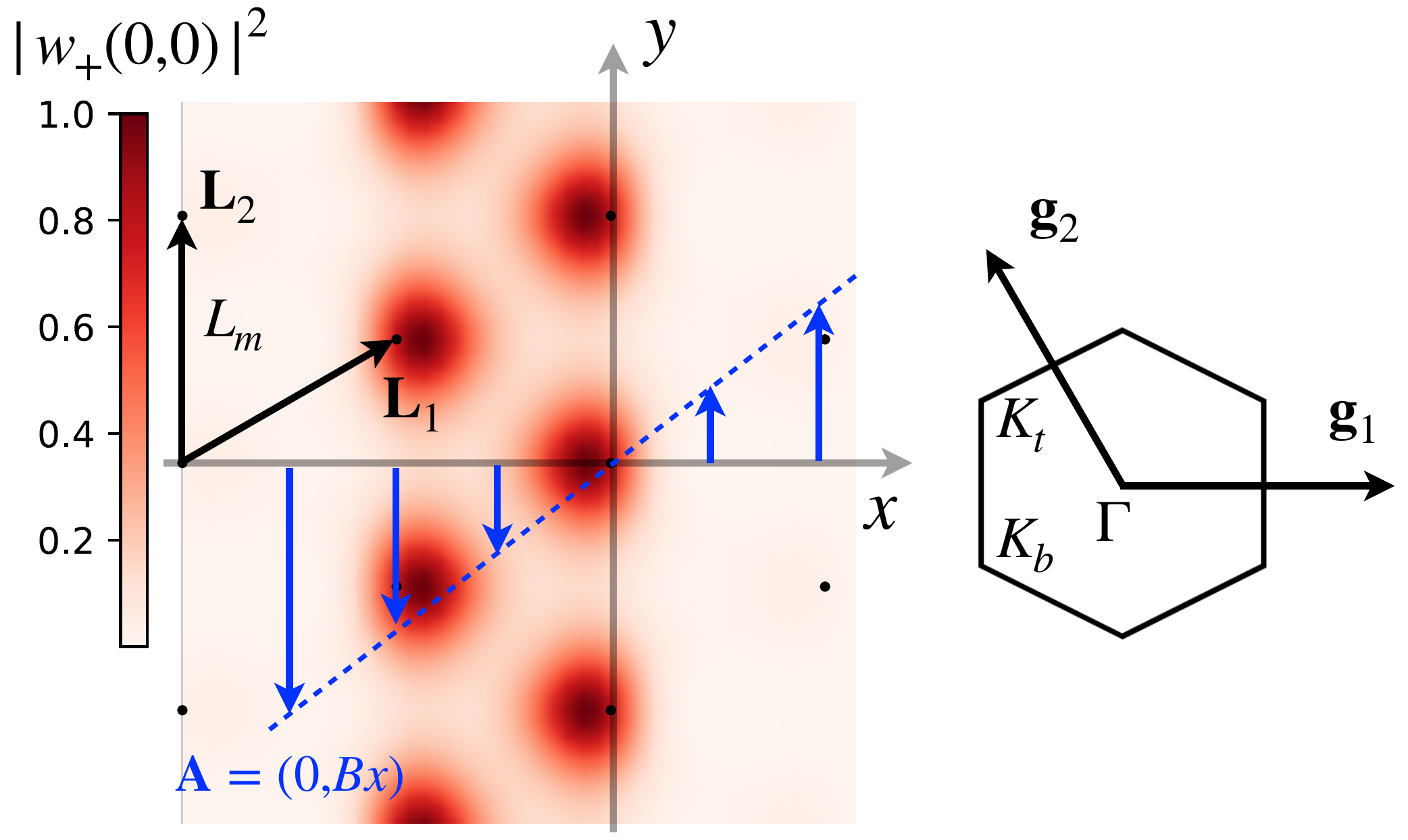}
    \caption{Left: illustrative real space probability density of a hybrid Wannier state $\ket{w_c(n_0,k_2\mathbf{g}_2)}$, with Chern index $c=+1$, $n_0=0$ and $k_2=0$, and the Landau gauge magnetic vector potential $\mathbf{A}=B x \hat{\by}$. Moir\'e unit cell primitive vectors are $\mathbf{L}_{1,2}$.  Right: moir\'e Brillouin zone and reciprocal lattice vectors $\mathbf{g}_{1,2}$. $K_{t,b}$ denote the Dirac point from the top and bottom layers of the twisted bilayer graphene.}
    \label{fig:schematic}
\end{figure}

\begin{figure}
    \centering
    \includegraphics[width=\linewidth]{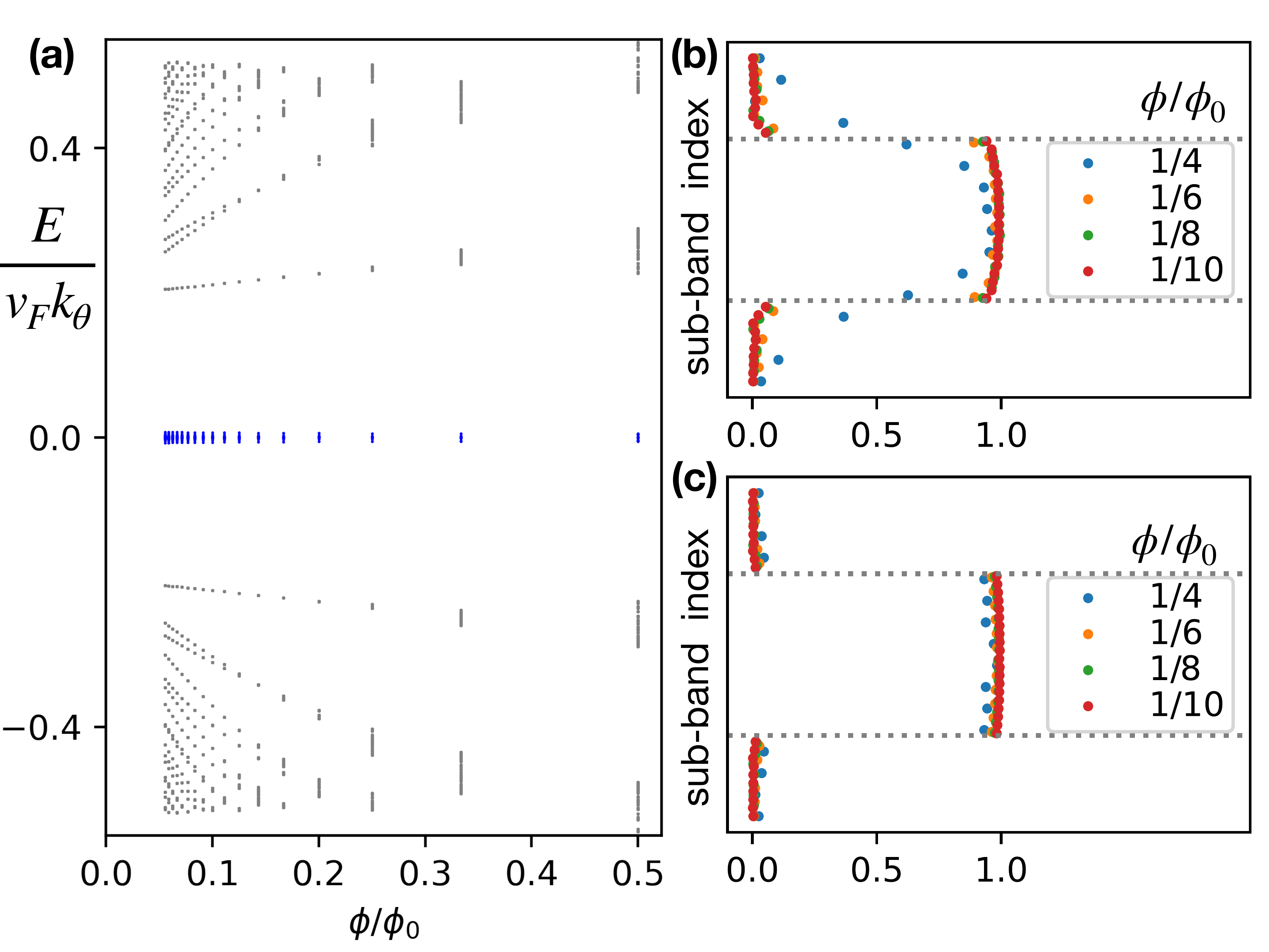}
    \caption{(a) Hofstadter spectrum for the non-interacting BM Hamiltonian $\hat{H}_{BM}^{\bK}(p_x,p_y-\frac{eB}{c}x)$ calculated using Landau level basis at magic angle ${w_1}/{v_Fk_\theta}=0.586$ and ${w_0}/{w_1}=0.7$. The horizontal axes in (b) and (c) are the overlaps between the $\bB=0$ narrow bands hybrid Wannier states projected onto the $\bB\neq 0$ representation of the magnetic translation group $|V_a\rangle$, and the exact magnetic subband states at various energies obtained using the LL basis $\sum_a|\langle \Psi_{\bK,m}(k_1,k_2)|V_{\bK,a}(k_1,k_2)\rangle|^2$; the vertical axes are the sub-band index $m$. The states in between the dashed lines belong to the $\bB\neq 0$ narrow bands shown as blue in (a), demonstrating that at low $\bB$, $|V_a\rangle$ have support almost exclusively within the $\bB\neq 0$ narrow band Hilbert space. ${w_0}/{w_1}=0.7$ in (b) and (c) is at the chiral limit ${w_0}/{w_1}=0$.}
    \label{fig:BMhoff and projector}
\end{figure}

The new method introduced here avoids the above mentioned difficulties. We illustrate it at low $\bB$, but the method is readily generalizable to vicinity of simple fractions $p/q$. Thus, we first solve the $\bB=0$ problem using standard (efficient) methods and find the hybrid Wannier states for the $\bB=0$ narrow bands \cite{Rui2011,Zhida2019,Kang2020a,Soejima2020}. Such states are exponentially localized in one direction and Bloch extended in another, say $y$-direction \cite{Kang2020a} (see Fig.~\ref{fig:schematic}). We stress that even if the band is topologically non-trivial, there is no obstruction to 1D exponential localization. The key insight is that at $\bB\neq 0$, for the hybrid Wannier state (WS) centered at and near the origin, the Landau gauge vector potential ${\bf A}=Bx \hat{{\bf y}}$ can be treated perturbatively, because the region in real space where $\bA$ is large gets suppressed by the exponential localization of the hybrid WS (see Fig.~\ref{fig:schematic}). Moreover, the discrete translation symmetry along the $y$-direction used in constructing the hybrid WSs is preserved by such $\bA$.
Next, we generate the rest of the basis by projecting the hybrid WSs centered at and near the origin onto a representation of the magnetic translation group. This gives us two quantum numbers, $k_1\in [0,1)$ and $k_2\in [0,1/q)$, associated with magnetic translations by two non-collinear vectors $\bL_1$ and $q\bL_2$ (Fig.~\ref{fig:schematic}). States with different $k_1$ and $k_2$ are then guaranteed to be orthogonal.
Because in the original ($\bB=0$) Brillouin zone $k_2$ belonged to a larger range $[0,1)$, we generate $q$ states for each starting hybrid WS at the same $k_1\in [0,1)$ and $k_2\in [0,1/q)$ when $\bB\neq 0$. Thus, for each $\bB=0$ narrow band (of which there are two per valley and spin in MATBG) and for each hybrid WS center described by a discrete index $n_0$, we have $q$ states. The resulting states at the same $k_1\in [0,1)$ and $k_2\in [0,1/q)$ then typically are not orthogonal, but by adjusting the range of $n_0$, the set of states can be readily made overcomplete and thus span the $\bB\neq 0$ narrow band Hilbert space. A simple procedure involving diagonalization of the overlap matrix and keeping the $2q$ largest overlap eigenvalues (per spin and valley) is then applied to obtain $2q$ orthogonal states within the MATBG narrow bands at $\bB\neq 0$. For MATBG and at low $\bB$ we find that the largest $2q$ overlap eigenvalues are clearly separated by a gap from the remaining small eigenvalues, and that the $2q$ orthogonal states have an almost perfect support by the $\bB\neq 0$ narrow bands only (see Fig.~\ref{fig:BMhoff and projector}b and \ref{fig:BMhoff and projector}c). 

\begin{figure}
    \centering
    \includegraphics[width=\linewidth]{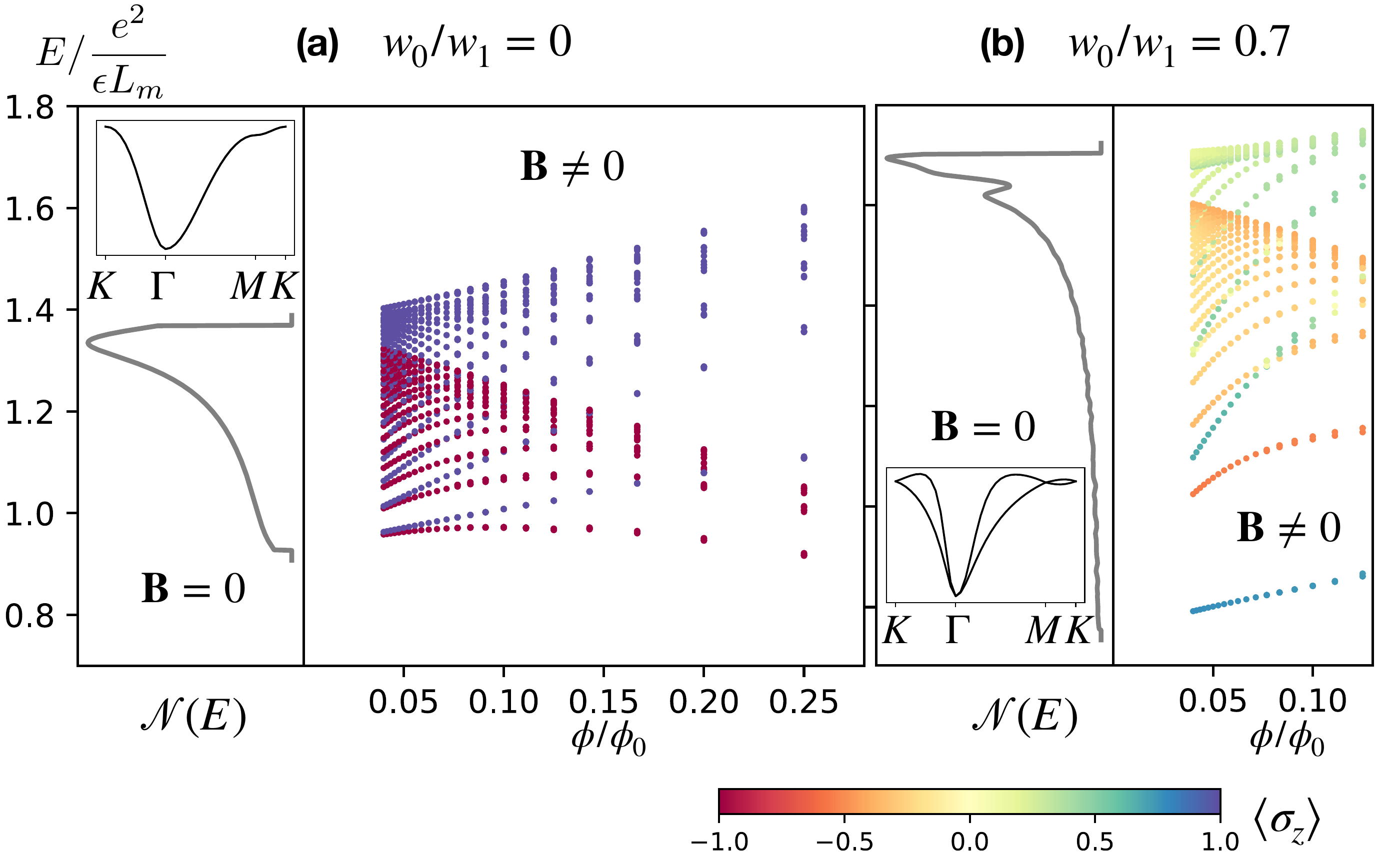}
    \caption{Landau level spectrum at magic angle in the strong coupling limit for (a) $w_0/w_1=0$ and (b) $w_0/w_1=0.7$. The gray lines denote the respective density of states $\mathcal{N}(E)$ at $\bB=0$, and the color of each magnetic sub-band denotes the average value of its sublattice polarization. Blue (red) denotes purely A (B) sublattice polarization.}
    \label{fig:strong_coupling}
\end{figure}

If we use this method on a topologically trivial narrow band, then a single value of $n_0=0$ (i.e. only the hybrid WS centered at the origin) is sufficient and none of the overlap eigenvalues become small even when $\phi=\phi_0$.
On the other hand, for the topologically non-trivial narrow bands of MATBG, we need to keep at least two starting states with $n_0=0$ and $n_0=\pm 1$ (for either sign) in order to obtain complete orthogonal basis spanning the $\bB\neq 0$ narrow bands. This is a direct consequence of the non-trivial topology of the $\bB=0$ narrow band Hilbert space, spanned by a band with Chern number $+1$ and a band with Chern number $-1$, one of which is then deficient by $p$ anomalous sub-bands while the other has an excess of $p$ sub-bands when $\bB\neq 0$ \footnote{In Supplemental Material we provide a derivation of $p$ anomalous magnetic sub-bands at flux $p/q$ in the chiral limit, and provide numerical evidence both in and away from the chiral limit.} \cite{Popov2021,Sheffer2021}. We confirm this by studying the sublattice polarization of the resulting states in Fig.~\ref{fig:strong_coupling} and analytical arguments in the chiral limit presented in SM.

\begin{figure}
    \centering
    \includegraphics[width=\linewidth]{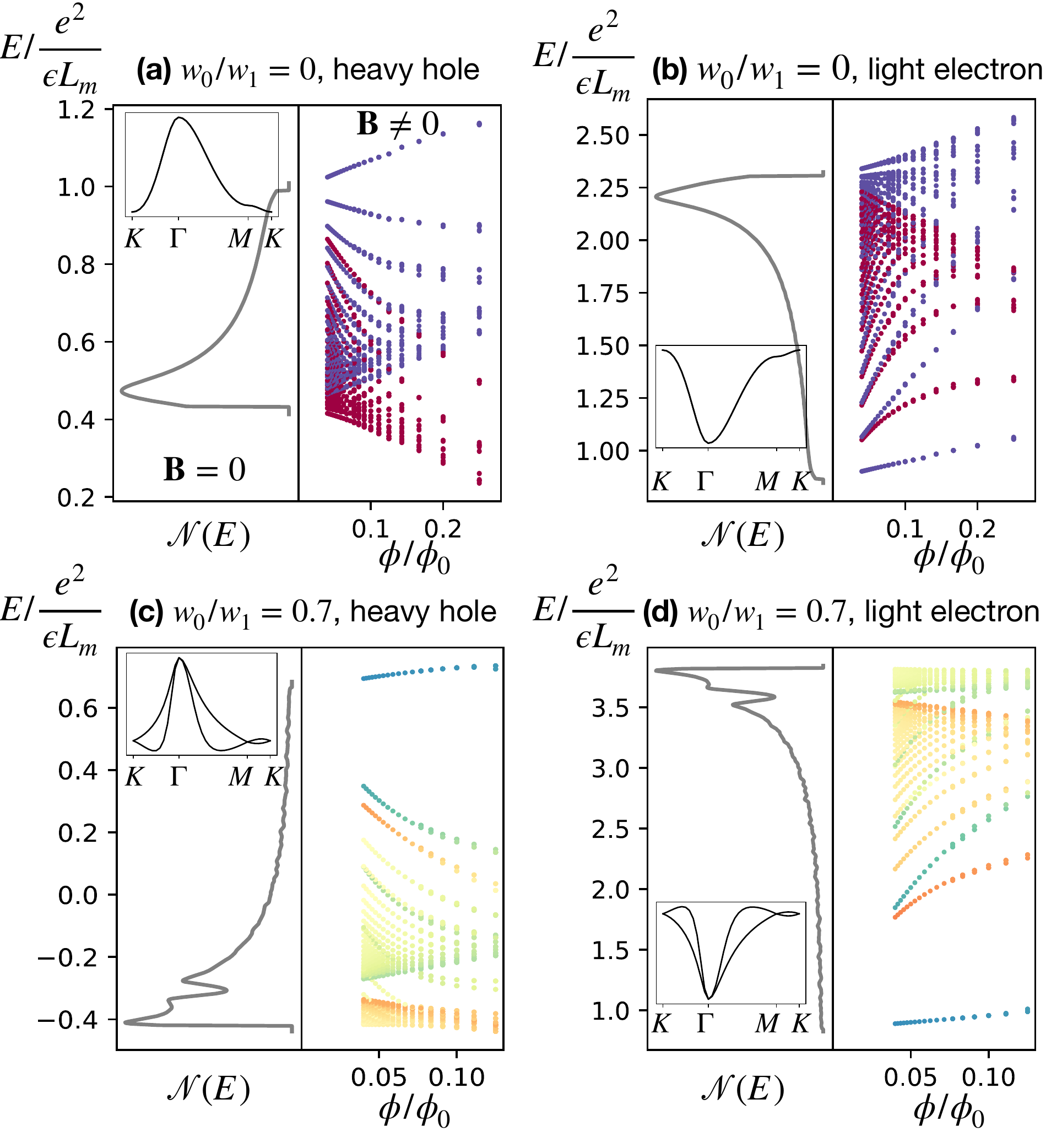}
    \caption{Landau level spectrum of charge $\pm 1$ excitations at $\nu=2$ at magic angle in the strong coupling limit. Heavy hole (a,c) and light electron (b,d) excitations for $w_0/w_1=0$ and $w_0/w_1=0.7$ respectively \cite{TBGV2020,Kang2021}. The color scale for sublattice polarization is the same as in Fig.~\ref{fig:strong_coupling}.}
    \label{fig:nu2hofstadter}
\end{figure}

Our new basis can now be readily applied to finding the $\bB\neq 0$ single electron or single hole excitation spectra in the strong coupling problem by using the method introduced in Refs.~\cite{Kang2020b,TBGV2020}.
Note that even at $\bB\neq0$, the 2-fold rotation about the out-of-plane axis $C_2$, the particle-hole $P$ \cite{Zhida2019,Hejazi2019} and the valley $U(1)$ conservation symmetries of the BM Hamiltonian are preserved at any $w_0/w_1$; the time reversal symmetry $T$ is of course broken by $\bB$. Therefore, $C_2P$ guarantees that if $\Psi_{\bK,m,k_1,k_2}(\br)$ is an eigenstate of $\hat{H}^{\bK}_{BM}\left(p_x,p_y-\frac{eB}{c}x\right)$ defined via Eq.~(\ref{eq:HBM}) below with an eigenvalue $E_{\bK,m,k_1,k_2}$, then $-i\mu_y\sigma_x e^{-i\bq_1\cdot\br}\Psi_{\bK,m,k_1,k_2}(\br)$ is an opposite valley eigenstate of  $\hat{H}^{\bK'}_{BM}\left(p_x,p_y-\frac{eB}{c}x\right)$ with an eigenvalue $-E_{\bK,m,k_1,k_2}$. The Pauli matrices $\sigma$ and $\mu$ act in the sublattice and layer spaces, respectively.
Eliminating the remote magnetic sub-bands using the RG procedure introduced in Ref.~\cite{Kang2020b} therefore still results in the residual Coulomb interaction projected onto the $\bB\neq 0$ narrow band Hilbert space to be of the form expressed in Eq.~(\ref{eq:Hint}). Moreover, ignoring the Zeeman effect, $C_2P$ guarantees that the spin valley $U(4)$ symmetry \cite{Kang2019,Bultinck2020b,TBGIII2020,Jonah2021} is still present even at $\bB\neq 0$. We can therefore follow the double commutator method outlined in Refs.~\cite{Kang2020b,TBGV2020} in order to find the spectrum of the single particle or single hole excitations at $\bB\neq 0$. The solutions of the Eq.~(\ref{eq:double commutator}) for two-gate screened Coulomb interaction, $V_{\bq}=\frac{2\pi e^2}{\eps |\bq|}\tanh\left(\frac{|\bq|\xi}{2}\right)$, with the gate separation $\xi=L_m$ are shown in the Fig.~\ref{fig:strong_coupling} for the charge neutral point (CNP, i.e. $\nu=0$), together with their $\bB=0$ density of states. The results at $|\nu|=2$ for the heavy and light mass sides are shown in the Fig.\ref{fig:nu2hofstadter}.
Below we provide details of the calculations which lead to the stated results. 

To obtain the narrow band Hilbert space, we start by considering the BM model at $\bB\neq 0$ in Landau gauge
$\hat{H}^{\bK}_{BM}\left(p_x,p_y-\frac{eB}{c}x\right)$ 
where at the valley $\bK$
\begin{eqnarray}\label{eq:HBM}
\hat{H}^\bK_{BM}(p_x,p_y) &=&\left(\begin{array}{cc} v_F\sigma\cdot\bp & T(\br)e^{i\bq_1\cdot\br} \\
e^{-i\bq_1\cdot\br} T^\dagger(\br) & v_F\sigma\cdot\left(\bp+\hbar\bq_1\right)
\end{array}\right).
\end{eqnarray}
The Hamiltonian in valley $\bK'$ can be obtained by first applying time reversal to $\hat{H}^\bK_{BM}(p_x,p_y)$ followed by the minimal substitution $p_y\rightarrow p_y-\frac{eB}{c}x$.
The Pauli matrices $\sigma$ act in the sublattice space \footnote{We ignore the small rotation of $\sigma$ matrices which was shown to lead to negligible effects on the narrow band Hilbert space.}.
The interlayer hopping functions are
$T(\br)=\sum_{j=1}^3 T_j e^{-i\bq_j\cdot\br}$
where $\bq_1=k_\theta(0,-1)$, $\bq_{2,3}=k_\theta\left(\pm\frac{\sqrt{3}}{2},\frac{1}{2}\right)$, $k_\theta=\frac{8\pi}{3a_0}\sin\frac{\theta}{2}=4\pi/(3L_m)$, $a_0 \approx 0.246$nm, $L_m$ is the period of the moir\'e lattice,
and $
T_{j+1}=w_01_2+w_1\left(\cos\left(\frac{2\pi}{3}j\right)\sigma_x+\sin\left(\frac{2\pi}{3}j\right)\sigma_y\right)$,
where $1_n$ is an $n\times n$ unit matrix.
At $\bB=0$, $\hat{H}^\bK_{BM}$ is invariant under discrete translations by any integer multiple of $\bL_1=L_m\left(\frac{\sqrt{3}}{2},\frac{1}{2}\right)$ and $\bL_2=L_m\left(0,1\right)$.
At $\bB\neq 0$ and in the chosen gauge $\hat{H}_{BM}^\bK$ is still invariant under the translation by $\bL_2$, but a translation by $\bL_1$ needs to be accompanied by a gauge transformation,
\begin{eqnarray}
\psi(\br)\rightarrow \hat{t}_{\bL_1}\psi(\br)=e^{i\frac{eB}{\hbar c}L_{1x}y}\psi(\br-\bL_1).
\end{eqnarray}
Thus, if $\psi(\br)$ is an eigenstate then so is $e^{i\frac{eB}{\hbar c}L_{1x}y}\psi(\br-\bL_1)$.
Translations by $\bL_2$ are generated by $\hat{t}_{\bL_2}\psi(\br)=\psi(\br-\bL_2)$.
Then $\hat{t}_{\bL_2}\hat{t}_{\bL_1}=e^{-2\pi i \phi/\phi_0}\hat{t}_{\bL_1}\hat{t}_{\bL_2}$,
where $\phi_0=\frac{hc}{e}$ and $\phi=B L_{1x}L_m$.
If $\phi/\phi_0=p/q$, with $p$ and $q$ relatively prime integers, then
$ \left[\hat{t}^q_{\bL_2},\hat{t}_{\bL_1}\right]=0.$

The $\bB=0$ hybrid WSs, $|w_{\pm}(n,k\bg_2)\rangle$, can be chosen to be eigenstates of the periodic position operator $\hat{O}=\hat{P} e^{-i\frac{1}{N_1}\bg_1\cdot \br} \hat{P}$, projected using $\hat{P}$ onto the $\bB=0$ narrow band Hilbert space studied (for details see Ref.\cite{Kang2020a}); here $N_1$ is a large integer.
The eigenvalues $e^{-2\pi i \frac{1}{N_1}\left(n+\langle x_{\pm}\rangle_k/|\bL_1|\right)}$ give the Wilson loops \cite{Rui2011,Zhida2019,Kang2020a,Soejima2020} for the Chern $+1$ and Chern $-1$ hybrid WSs.
These states are localized along $\bL_1$ and Bloch extended along $\bL_2$, as illustrated in the Fig.~\ref{fig:schematic}
As shown in Ref.~\cite{Kang2020a}, they satisfy,
\begin{eqnarray}
\hat{t}_{\bL_1}|w_{\pm}(n,k_2\bg_2)\rangle &=& e^{i\frac{eB}{\hbar c}L_{1x}y}|w_{\pm}(n+1,k_2\bg_2)\rangle \\
\hat{t}_{\bL_2}|w_{\pm}(n,k_2\bg_2)\rangle &=& e^{-2\pi i k_2}|w_{\pm}(n,k_2\bg_2)\rangle.
\end{eqnarray}
We construct our basis for the narrow band at $\bB\neq0$ by projecting $|w_{\pm}(n_0,k_2\bg_2)\rangle$ onto representation of the magnetic translation group (MTG). We include in our set a range of $n_0$'s near $0$ as
\begin{equation}
|W_{\pm}(k_1,k_2; n_0)\rangle =\frac{1}{\sqrt{N}}\sum_{s=-\infty}^\infty e^{2\pi i s k_1} \hat{t}^s_{\bL_1} |w_{\pm}(n_0,k_2\bg_2)\rangle,
\end{equation}
with normalization factor $N$ and for $k_1\in [0,1)$ and temporarily letting $k_2\in [0,1)$. 
The results in Figs.\ref{fig:BMhoff and projector}b, \ref{fig:BMhoff and projector}c and \ref{fig:strong_coupling} include $n_0=0$ and $1$.
Note that $|W_{\pm}(k_1,k_2;n_0)\rangle$ are simultaneous eigenstates of $\hat{t}_{\bL_1}$ and $\hat{t}^q_{\bL_2}$ with eigenvalues $e^{-2\pi i k_1}$ and $e^{-2\pi i q k_2}$, respectively.
Thus the $q\bL_2$ translations break up the $k_2$ domain into $q$ pieces of equal width $1/q$. Therefore, we let $\left|W_{\pm}\left(k_1,k_2+{l}/{q}; n_0\right)\right\rangle$, permanently fix $k_2\in[0,1/q)$, and let $l=0,1,\ldots q-1$.
For different values of $k_1$ and $k_2$ in their respective domains $\left|W_{\pm}\left(k_1,k_2+{l}/{q}; n_0\right)\right\rangle$'s are orthogonal because they have different eigenvalues under $\hat{t}_{\bL_1}$ and $\hat{t}^q_{\bL_2}$.
For the same $k_1$ and $k_2$, but different $l$ (and different $n_0$) the states $\left|W_{\pm}\left(k_1,k_2+{l}/{q}; n_0\right)\right\rangle$'s are in general not orthogonal.
To orthogonalize them we diagonalize the overlap matrix $M_{ab}=\langle W_a|W_b\rangle=\left(U^\dagger D U\right)_{ab}$ where $D$ is diagonal. In the above  we combined $l$, the Chern number index $c=\pm$, and $n_0$ into a single index $a$ for each $k_1$ and $k_2$, whose dependence we temporarily suppress. Then we let
\begin{equation}
|V_a\rangle = \sum_b|W_b\rangle U^\dagger_{ba}\frac{1}{\sqrt{D_{a}}}
\end{equation}
where $b$ runs over all the indices but $a$ runs only over the $2q$ largest eigenvalues $D_a$.
As demonstrated in Fig.~\ref{fig:BMhoff and projector}b and Fig.~\ref{fig:BMhoff and projector}c, at low $\bB$, the $2q$ orthogonal states $|V_{a}(k_1,k_2)\rangle$ at each $k_1$ and $k_2$ now form the basis spanning almost exclusively only the $\bB\neq 0$ narrow bands. At larger $\bB$, we find a spillover into the remote bands; for the results presented the spillover is negligible.

Next, we turn to the excitations in the strong coupling limit.
In this case the Hamiltonian consists of only the interaction $V(\br-\br')$ projected onto the $\bB\neq 0$ narrow band basis. As described earlier, the $C_2P$ symmetry guarantees that the dominant term takes the form
\begin{equation}\label{eq:Hint}
\hat{H}_{int}=\frac{1}{2}\int d\br \int d\br' V(\br-\br')\delta\rho(\br)\delta\rho(\br'),
\end{equation}
where $\delta \rho(\br)=\rho(\br)-\bar{\rho}(\br)$. Restoring the indices on our $\bB\neq 0$ narrow band basis functions $\langle \br|V_a\rangle$, 
the projected density operator is $\rho(\br)=$
\begin{equation}
\sum_{k_1,k_2,a}\sum_{k'_1,k'_2,a'}V^\dagger_{\bK, a}(k_1,k_2;\br)V_{\bK, a'}(k'_1,k'_2;\br)\frak{d}^\dagger_{a,k_1,k_2} \frak{d}_{a',k'_1,k'_2}.
\end{equation}
We arranged the fermion creation operators with $2q$ discrete quantum numbers $a$ and momentum $k_1$,$k_2$ into 4-component ``spinor'',
$\frak{d}^\dagger_{a,k_1,k_2}=\left(d^\dagger_{\uparrow, \bK;a,k_1,k_2},d^\dagger_{\downarrow,\bK;a,k_1,k_2},d^\dagger_{\uparrow, \bK';C_2P[a,k_1,k_2]},d^\dagger_{\downarrow,\bK';C_2P[a,k_1,k_2]}\right)$. 
The $U(4)$ manifold can be generated from a valley polarized state, which is an eigenstate of $\rho(\br)$ with the eigenvalue equal to $\bar{\rho}(\br)$ at CNP, where it takes the form, say, $|\Phi_{\nu=0}\rangle=\prod_{a,k_1,k_2,\sigma=\uparrow,\downarrow}d^\dagger_{\sigma,\bK;a,k_1,k_2}|0\rangle$.
Excitations can be created using an operator $X$ (see Ref.~\cite{Kang2020b}) and their strong coupling eigenenergies can be read off from the equation
\begin{eqnarray}\label{eq:double commutator}
E X|\Phi_\nu\rangle &=& \frac{1}{2}\int d^2\br d^2\br' V(\br-\br')\left[\rho(\br),\left[\rho(\br'),X\right]\right]|\Phi_\nu\rangle\nonumber\\
&+&\int d^2\br d^2\br' V(\br-\br')\left[\rho(\br),X\right]\delta\bar{\rho}(\br')|\Phi_\nu\rangle,
\end{eqnarray}
where we extended the result to include $\nu=\pm 2$ fillings \cite{TBGV2020,Kang2021}; the valley polarized states $|\Phi_\nu\rangle$ are eigenstates of $\delta\rho(\br)$ with an eigenvalue $\delta\bar{\rho}(\br)$ \cite{TBGIV2020}.  
The eigenenergies of the strong coupling single particle or single hole excitations can now be determined from diagonalizing a $2q\times 2q$ matrix for each $k_1$ and $k_2$.
Their degeneracy is determined by considering the action of $X$ on $|\Phi_\nu\rangle$.

The resulting spectra at CNP are shown in the right panel of Fig.~\ref{fig:strong_coupling}a for the chiral limit $w_0/w_1=0$ and the right panel of Fig.~\ref{fig:strong_coupling}b for $w_0/w_1=0.7$; the spectra at $\nu=2$ are shown in Fig.~\ref{fig:nu2hofstadter}. 
We clearly see that despite being at strong coupling the excitations' spectra are Landau quantized in $\bB\neq 0$. In the chiral limit (Fig.~\ref{fig:strong_coupling}a), the degeneracy of the low lying excitations limits to $4$ at low $\bB$ due to spin and sublattice degrees of freedom, the latter taking on values $\pm 1$ as marked by the blue and red colors. Because they originate from $\bB=0$ Chern bands with opposite total Chern numbers, the $B$ sublattice sector has $q-1$ sub-bands while the $A$ sublattice sector has $q+1$ sub-bands for the $1/q$ sequence shown. Note that at small $\bB$ there is a small splitting between the low lying opposite sublattice polarized strong coupling sub-bands due to broken $C_2T$ symmetry and that this splitting increases with increasing $\bB$. A similar conclusion has been reached in a recent theoretical work~\cite{Jonah2021}, which reported energy splitting of the charge-$\pm1$ excitations at full flux $\phi/\phi_0=1$. Also note the opposite evolution of the sub-bands emanating from the $\bB=0$ van Hove singularities. Many of the features are reproduced at $w_0/w_1=0.7$, except the smaller mean value of the sublattice polarization (as marked by the color scheme), and larger splitting between the low lying magnetic sub-bands. Interestingly, the sizable splitting between the light fermion LLs seen for $w_0/w_1=0.7$ in Figs.~\ref{fig:strong_coupling}b and \ref{fig:nu2hofstadter}d even at small $\phi/\phi_0$ would give rise to prominent LL filling factors $|\nu_{LL}|=0,2$ at CNP, and $\nu_{LL}=0,1$ on the light mass side of $\nu=2$, as observed in Ref.~\cite{Yacoby2021} without invoking moir\'e translational symmetry breaking. 

Published STM spectroscopy data \cite{Nuckolls2020} at $\bB\neq 0$ show only results from regions of various devices with the values of heterostrain $0.1\%-0.4\%$. It is known that even such small values of strain dramatically increase the non-interacting narrow band width \cite{ZhenBi2019,Parker2021}, making the kinetic energy comparable or larger than the Coulomb interaction scale $e^2/\eps L_m$, and stabilizing energetically proximate nematic state \cite{Shang2021,Kang2020a,Parker2021}. Therefore, the available STM data~\cite{Nuckolls2020} at $\bB\neq0$ may not be in the limit dominated by the Coulomb interaction complicating the direct comparison with the strong coupling result presented here. The spectroscopic measurements on magic angle devices at $\bB\neq 0$ with negligible strain would therefore be highly desirable. 

The method presented here is general, and can be used to find the Hofstadter spectrum at larger $\phi/\phi_0$ by starting with simple fractions $1/\bar{q}$, where $\bar{q}$ is a small integer such as $1$,$2$ or $3$ and where the LL based calculation is manageable, building the hybrid WSs for the $2\bar{q}$ Hofstadter bands, and then projecting onto the representation of the magnetic translation group for $\phi/\phi_0$ away from $1/\bar{q}$. Such generalizations, as well as strain effects will be presented in future work.
 
\begin{acknowledgments}
We thank B. Andrei Bernevig, Jonah Herzog-Arbeitman, Jian Kang for helpful discussions. X.W. acknowledges financial support from National 
MagLab through Dirac fellowship, which is funded by the National Science Foundation (Grant No. DMR-1644779) and the state of Florida. O.V. was supported by NSF Grant No. DMR-1916958. 
\end{acknowledgments}

\bibliographystyle{apsrev4-2}
\bibliography{references}

\appendix
\widetext
\begin{center}
\textbf{\large Supplemental Materials for ``Narrow bands in magnetic field and strong-coupling Hofstadter spectra"}
\end{center}
\setcounter{equation}{0}
\setcounter{figure}{0}
\setcounter{table}{0}
\setcounter{page}{1}
\makeatletter
\renewcommand{\theequation}{S\arabic{equation}}
\renewcommand{\thefigure}{S\arabic{figure}}
\renewcommand{\bibnumfmt}[1]{[#1]}
\renewcommand{\citenumfont}[1]{#1}
\section{Index theorem and analytical results in the chiral limit $w_0/w_1=0$.}
The non-interacting results in this section appeared previously in Refs.\cite{Popov2021,Sheffer2021}; the strong coupling results relevant for the main text are new. We include the re-derivation of the former for convenience.

Let us adopt the symmetric gauge $\bA=\frac{1}{2}B\left(-y,x,0\right)$.
\begin{eqnarray} 
\hat{H}^\bK_{BM}\left(\bp-\frac{e}{c}\bA \right)&=&\left(\begin{array}{cc}
v_F \sigma\cdot\left(\bp-\frac{e}{c}\bA\right) & T(\br)e^{i\bq_1\cdot\br} \label{eq:HBM}\\
e^{-i\bq_1\cdot\br}T^\dagger(\br) &  v_F \sigma\cdot\left(\bp+\bq_1-\frac{e}{c}\bA\right)
\end{array}\right);\\
v_F \sigma\cdot\left(\bp+\bq_1-\frac{e}{c}\bA\right)
&=&
\hbar v_F\left(\begin{array}{cc}0 & \frac{1}{i}\frac{\partial}{\partial x}-\frac{\partial}{\partial y}+ik_\theta+i\frac{eB}{2\hbar c}(x-iy)\\
\frac{1}{i}\frac{\partial}{\partial x}+\frac{\partial}{\partial y}-ik_\theta-i\frac{eB}{2\hbar c}(x+iy) & 0
\end{array}\right)\\
&=&
\hbar v_F\left(\begin{array}{cc}0 & \frac{2}{i}\frac{\partial}{\partial z}+ik_\theta+i\frac{eB}{2\hbar c}\bar{z}\\
\frac{2}{i}\frac{\partial}{\partial \bar{z}}-ik_\theta-i\frac{eB}{2\hbar c}z & 0
\end{array}\right)
\end{eqnarray}
where $z=x+iy$, $\bar{z}=x-iy$, $2\frac{\partial}{\partial z}=\frac{\partial}{\partial x}-i\frac{\partial}{\partial y}$ and $2\frac{\partial}{\partial \bar{z}}=\frac{\partial}{\partial x}+i\frac{\partial}{\partial y}$.
In the chiral limit,
\begin{eqnarray}
T(\br)&=&w_1\left(\begin{array}{cc}0 & e^{-i\bq_1\cdot\br}+ e^{-i\frac{2\pi}{3}} e^{-i\bq_2\cdot\br}
+ e^{i\frac{2\pi}{3}} e^{-i\bq_3\cdot\br}
 \\
e^{-i\bq_1\cdot\br}+ e^{i\frac{2\pi}{3}} e^{-i\bq_2\cdot\br}
+ e^{-i\frac{2\pi}{3}} e^{-i\bq_3\cdot\br}
 & 0
\end{array}\right).
\end{eqnarray}
In the above, $\hat{H}^\bK_{BM}$ acts on $\left(A_{top},B_{top},A_{bot},B_{bot}\right)$. Consider the unitary transformation after which we have $\hat{H}^\bK_{BM}$ act on $\left(A_{top},A_{bot},B_{top},B_{bot}\right)$ as
\begin{eqnarray}
\hat{H}^\bK_{BM}&\rightarrow&\left(\begin{array}{cc} 0 & \mathcal{D}^\dagger \\ \mathcal{D} & 0\end{array}\right),\;\;
\mathcal{D}=
\left(\begin{array}{cc}
-i\hbar v_F \left(2\frac{\partial}{\partial \bar{z}}+\frac{z}{2\ell^2}\right) & w_1 U(\br)e^{i\bq_1\cdot\br} \\
w_1 U(-\br)e^{-i\bq_1\cdot\br} & -i\hbar v_F \left(2\frac{\partial}{\partial \bar{z}}+k_\theta+\frac{z}{2\ell^2}\right)
 \end{array}\right).
\end{eqnarray}
Here $\ell^2=\hbar c/(eB)$ and $U(\br)=e^{-i\bq_1\cdot\br}+ e^{i\frac{2\pi}{3}} e^{-i\bq_2\cdot\br}
+ e^{-i\frac{2\pi}{3}} e^{-i\bq_3\cdot\br}$.
Therefore, any state of the form
\begin{eqnarray}\label{eqn:zero modes sym gauge}
e^{-\frac{1}{4\ell^2}\bar{z}z}f(z) \left(\begin{array}{c} \Psi^{chiral}_{\bK_m}(\br) \\ 0 \end{array}\right)
\end{eqnarray}
and
\begin{eqnarray}
e^{-\frac{1}{4\ell^2}\bar{z}z}f(z) \left(\begin{array}{c} \Psi^{chiral}_{\bK'_m}(\br) \\ 0 \end{array}\right)
\end{eqnarray}
is a normalizable zero energy solution for an analytic $f(z)$, because $\Psi^{chiral}_{\bK_m}(\br)$ and $\Psi^{chiral}_{\bK'_m}(\br)$ are the exact zero energy states at $\bB=0$; such a $\bB=0$ state can always be found even away from the magic angle \cite{Tarnopolsky2019}.
Note that these states live entirely on the $A$-sublattice and that unlike in the $\bB=0$ (see Ref.\cite{Tarnopolsky2019}), there is no normalizable solution on the $B$-sublattice.

Now, $f(z)\in \left(1,z,z^2,\ldots,z^N\right)$, where $N+1=N_\phi$ is the degeneracy of the Landau level.
To show that the two states in Eq.(\ref{eqn:zero modes sym gauge}) are linearly independent we need to show that the equation
\begin{eqnarray}\label{eqn:linear indep 1}
\sum_{n=0}^N c_n z^n e^{-\frac{1}{4\ell^2}\bar{z}z} \left(\begin{array}{c} \Psi^{chiral}_{\bK_m}(\br) \\ 0 \end{array}\right)
+
\sum_{n=0}^N c'_n z^n e^{-\frac{1}{4\ell^2}\bar{z}z} \left(\begin{array}{c} \Psi^{chiral}_{\bK'_m}(\br) \\ 0 \end{array}\right)=0
\end{eqnarray}
has a solution for all $\br$ only if $c_n=c'_n=0$ for all $n$.
To do so, we note that even at $\bB\neq 0$, we have $C'_2T = \mu_x K(x\rightarrow -x)$ symmetry and the unitary $P$ symmetry, whose combination also changes the sign of $\hat{H}^\bK_{BM}$:
\begin{eqnarray}
PC'_2T: \mu_z \hat{H}^{\bK^*}_{BM}(x,-y)\mu_z=-\hat{H}^\bK_{BM}(x,y).
\end{eqnarray}
This means that the zero modes can be chosen to be eigenstates of $PC'_2T$.
To find out the parity of the zero modes at $\bK_m$ and $\bK'_{m}$, we note that at $\bB=0$ the $\bk$-points $\bK_m$ and $\bK'_m$ are related by the particle hole symmetry $P=i\mu_y e^{i\bq_1\cdot\br}\hat{\mathcal{I}}$ where $\hat{\mathcal{I}}\psi(\br)=\psi(-\br)$.
Therefore, we can choose $\Psi^{chiral}_{\bK'_m}(\br)=\hat{P}\Psi^{chiral}_{\bK_m}(\br)$.
We see that $P$ {\it anti-commutes} with $P C'_2T$ and does not change the sublattice. Therefore, the parity of $\Psi^{chiral}_{\bK_m}(\br)$ and $\Psi^{chiral}_{\bK'_m}(\br)=\hat{P}\Psi^{chiral}_{\bK_m}(\br)$ must be opposite under $P C'_2T$.

We can also see this explicitly from a perturbative ``tripod model'' solution of Ref.\cite{TBGI2020}.
Up to a normalization, the approximate zero mode of interest at $\bK_m$ is
\begin{eqnarray}
\Psi^{chiral}_{\bK_m}(\br)=\sum_\bg\left(\begin{array}{c}A^{top}_{n,\bg}(-\bq_1)\\ A^{bot}_{n,\bg}(-\bq_1)\end{array}\right)e^{i\bg\cdot\br}\approx
\left(\begin{array}{c} 1+e^{-i(\bg_1+\bg_2)\cdot\br}+e^{-i\bg_2\cdot\br}\\ -i\frac{\hbar v_F k_\theta}{w_1}\end{array}\right)
\end{eqnarray}
and at $\bK'_m$ it is
\begin{eqnarray}
\Psi^{chiral}_{\bK'_m}(\br)=\sum_\bg\left(\begin{array}{c}A^{top}_{n,\bg}(0)\\ A^{bot}_{n,\bg}(0)\end{array}\right)e^{i\bg\cdot\br}\approx
\left(\begin{array}{c} i\frac{\hbar v_F k_\theta}{w_1} \\ 1+e^{i(\bg_1+\bg_2)\cdot\br}+e^{i\bg_2\cdot\br}\end{array}\right)
\end{eqnarray}
where $\bg_1+\bg_2=k_\theta\left(\frac{\sqrt{3}}{2},\frac{3}{2}\right)$ and
$\bg_2=k_\theta(-\frac{\sqrt{3}}{2},\frac{3}{2})$; $k_\theta=\frac{4\pi}{3L_m}$.
We see that
\begin{eqnarray}
PC'_2T:\;\; \mu_z  \Psi^{* chiral}_{\bK_m}(x,-y)&=&\Psi^{chiral}_{\bK_m}(x,y)\\
\;\; \mu_z  \Psi^{* chiral}_{\bK'_m}(x,-y)&=&-\Psi^{chiral}_{\bK'_m}(x,y),
\end{eqnarray}
in other words, they have opposite parity under $PC'_2T$.

Because $PC'_2T$ is a linear operator, going back to the equation defining the linear independence, we can apply $PC'_2T$ to both sides of the Eq.(\ref{eqn:linear indep 1}) to find
\begin{eqnarray}\label{eqn:linear indep 2}
PC'_2T:\;\; \sum_{n=0}^N c_n z^n e^{-\frac{1}{4\ell^2}\bar{z}z} \left(\begin{array}{c} \Psi^{chiral}_{\bK_m}(\br) \\ 0 \end{array}\right)
-
\sum_{n=0}^N c'_n z^n e^{-\frac{1}{4\ell^2}\bar{z}z} \left(\begin{array}{c} \Psi^{chiral}_{\bK'_m}(\br) \\ 0 \end{array}\right)=0.
\end{eqnarray}
Adding and subtracting Eqs.(\ref{eqn:linear indep 1}) and (\ref{eqn:linear indep 2}) we find
\begin{eqnarray}\label{eqn:linear indep 3}
&&\sum_{n=0}^N c_n z^n e^{-\frac{1}{4\ell^2}\bar{z}z} \left(\begin{array}{c} \Psi^{chiral}_{\bK_m}(\br) \\ 0 \end{array}\right)=0,\\
&&\sum_{n=0}^N c'_n z^n e^{-\frac{1}{4\ell^2}\bar{z}z} \left(\begin{array}{c} \Psi^{chiral}_{\bK'_m}(\br) \\ 0 \end{array}\right)=0.
\end{eqnarray}
Since the spinors and the gaussian factors are non-zero, the above hold only if
\begin{eqnarray}\label{eqn:linear indep 3}
&&\sum_{n=0}^N c_n z^n =0,\\
&&\sum_{n=0}^N c'_n z^n =0.
\end{eqnarray}
But polynomials of different degrees are linearly independent, as can be seen by taking successive derivatives and showing that the only way these equations are satisfied for all $z$ is if each coefficient vanishes identically.

This proves that we have two Landau levels worth of zero modes in the chiral limit at $\bB\neq 0$ at a general twist angle.
This number is equivalent to having two exact zero modes for $k_1\in [0,1)$ and $k_2\in [0,\frac{p}{q})$.

Because in the chiral limit
\begin{equation}
\{\hat{H}^\bK_{BM},1_2\sigma_z\}=0
\end{equation}
and because the sublattice polarization eigevalues are $\pm1$, by the index theorem (see e.g. Refs.\cite{Sheffer2021,Vafek2006}) we have
\begin{equation}
\mbox{Tr}\left[\mathcal{P}1_2\sigma_z\right]=n_+-n_-,
\end{equation}
where $\mathcal{P}$ is the projector onto the narrow band Hilbert space, and where  $n_+$ is the number of zero energy modes with $1_2\sigma_z$ eigenvalue $+1$ and $n_-$ is the number of zero energy modes with $1_2\sigma_z$ eigenvalue $-1$.
But we found all of the zero energy modes and they are sublattice $A$ polarized. Therefore, for a given $k_1\in [0,1)$ and $k_2\in [0,1/q)$ we have $\mbox{Tr}\left[\mathcal{P}1_2\sigma_z\right]=2p$ for any finite $\bB$ and independent of the twist angle (as long as BM model applies).
Note that at $\bB=0$, $\mbox{Tr}\left[\mathcal{P}1_2\sigma_z\right]=0$ because for every $A$-sublattice polarized zero mode there is a $B$-sublattice polarized zero mode\cite{Tarnopolsky2019}. 
Therefore, $\mbox{Tr}\left[\mathcal{P}1_2\sigma_z\right]$ is discontinuous at $\bB=0$.

\begin{figure}
    \centering
    \includegraphics[width=0.7\linewidth]{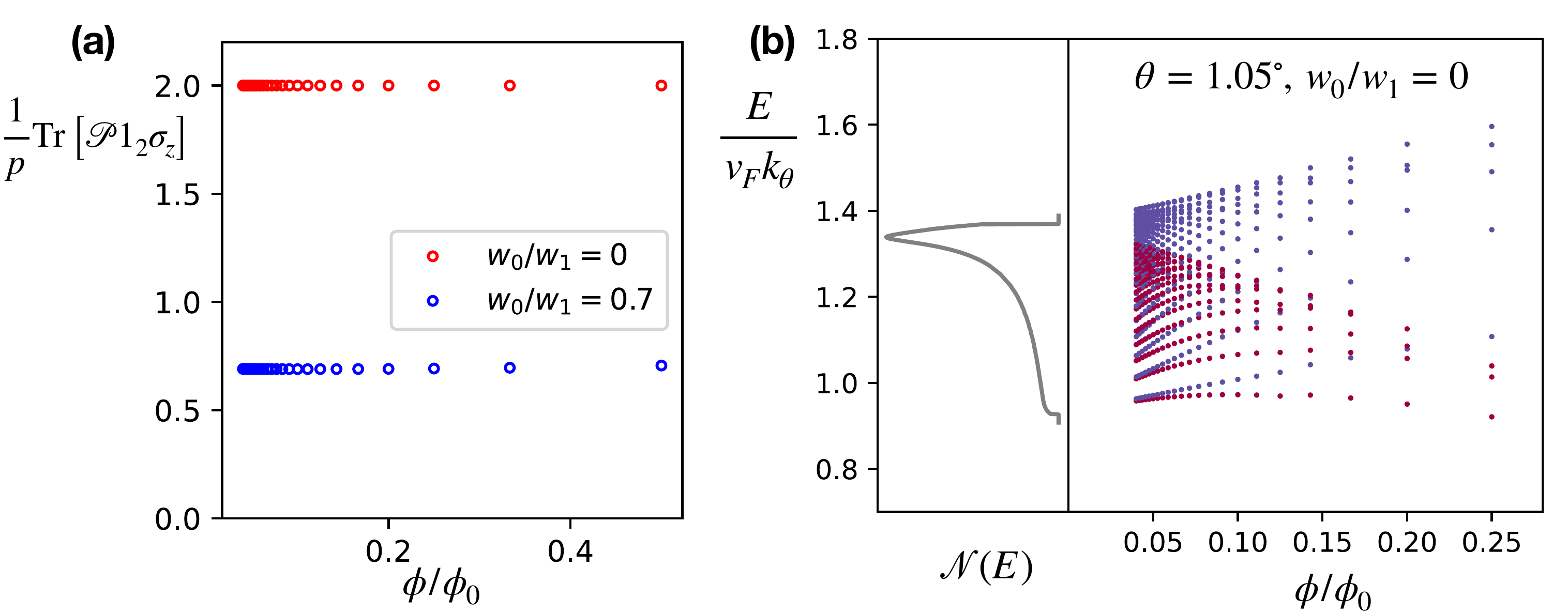}
    \caption{(a) Trace of sublattice polarization for the magic angle in chiral limit $w_0/w_1=0$ (red) and for $w_0/w_1=0.7$ (blue) as a function of magnetic flux $\phi$ through the moire unit cell in units of the flux quantum $\phi_0=hc/e$, where $\mathcal{P}$ is the projector onto the narrow bands in magnetic field. Due to the $C_2T$ symmetry at $\bB=0$, Tr$[\mathcal{P}1_2\sigma_z]$ must vanish at $\phi/\phi_0=0$; note the discontinuous jump of the trace of the projected sublattice polarization at non-zero $\phi/\phi_0$. (b) Spectrum of charge $\pm1$ excitations at CNP in the chiral limit for the $\phi/\phi_0=1/q$ sequence, for a single momentum $\bk=(0,0)$ in the magnetic Brillouin zone. There are $q+1$ A sublattice polarized (blue) states and $q-1$ B sublattice polarized (red) states.}
    \label{fig:figs1}
\end{figure}
We demonstrate the above result numerically in the Fig.~\ref{fig:figs1}(a). As seen, although the sublattice polarization is no longer perfect at $\frac{w_0}{w_1}\neq 0$, the discontinuity at $\bB=0$ persists.

In the strong coupling chiral limit the effective Hamiltonian for the single particle excitations {\it commutes} with $1_2\sigma_z$. Therefore, all of the strong coupling magnetic sub-bands can be chosen to be eigenstates of $1_2\sigma_z$ with eigenvalues $+1$ or $-1$. Since we have just proved that the narrow band Hilbert space onto which we projected the interaction has an extra $2p$ sublattice A polarized states at each $k_1\in [0,1)$ and $k_2\in [0,1/q)$, we must have $q+p$ sublattice $A$ polarized magnetic sub-bands and $q-p$ sublattice $B$ polarized sub-bands as demonstrated in Fig.~\ref{fig:figs1}(b), where for clarity we show the spectrum for a single value of $k_1$ and $k_2$ along the $\phi/\phi_0=1/q$ sequence.

Throughout the main text and the supplementary, we have defined the ``magic angle"  as the condition \cite{Tarnopolsky2019}:
\begin{equation}
    \frac{w_1}{v_Fk_\theta} \equiv 0.586,\  w_1=96.056\text{meV},\ \frac{\hbar v_F}{a}=2135.4\text{meV}.
\end{equation}
Here $a\approx 2.46 \AA$ is the graphene lattice constant. These parameter choices place the magic angle at $\theta=1.05^\circ$. 

\section{Hybrid Wannier approach in finite magnetic field and numerical procedure}
\subsection{Hybrid Wannier wavefunction representation of $\mathbf{B}=0$ narrow band}

In twisted bilayer graphene, hybrid Wannier wavefunctions have been constructed as basis states for describing the narrow band physics of the Bistritzer-MacDonald (BM) Hamiltonian at $\bB=0$. Unlike exponentially localized and symmetric Wannier states in both directions, there are no topological obstructions to constructing hybrid Wannier states, which are exponentially localized along one direction and Bloch extended along the other. Detailed discussions of hybrid Wannier states have been given in Refs.~\cite{Zhida2019,Kang2020a}. Here we merely outline the general procedure for constructing hybrid Wannier states. We begin by solving for the two narrow band energy eigenstates of the BM Hamiltonian per valley and spin (Eq.~(\ref{eq:HBM})), and construct them to also be eigenstates of the $C_2T$ operator with eigenvalue $+1$; this fixes their phase up to a sign. Next, the hybrid Wannier states $\ket{w_{c}(n_1,k_2\bg_2)}$ are constructed as eigenstates of the projected (periodic) position operator $\hat{P}e^{-i\frac{\bg_1}{N_1}\cdot\br}\hat{P}$, where $N_1$ is a large integer that discretizes the momentum space along the $\bg_1=\frac{4\pi}{\sqrt{3}L_m}(1,0)$ direction, and $\hat{P}$ is the projector onto the $\bB=0$ narrow bands. The hybrid Wannier states are labeled by the Chern number $c=\pm 1$, momentum $k_2\in[0,1)$ along the $\bg_2\equiv \frac{4\pi}{\sqrt{3}L_m}(-\frac{1}{2},\frac{\sqrt{3}}{2})$ direction, and index $n_1\in \mathbb{Z}$ denoting the exponential localization of the hybrid Wannier states near the real space strip $\br =(n_1L_{1x},y)$. They can be represented as a one-dimensional Fourier transform of the Chern Bloch states $\ket{\Psi_{c}(k_1\bg_1,k_2\bg_2)}$, smooth in $k_1$, as \cite{Kang2020a}:
\begin{equation}
    \ket{w_{c}(n_1,k_2\bg_2)}=\frac{1}{\sqrt{N_1}}\sum_{k_1=0}^{1-\frac{1}{N_1}}e^{-i2\pi k_1 n_1} \ket{\Psi_{c}(k_1\bg_1,k_2\bg_2)},
\end{equation}
where $k_1,k_2\in [0,1)$. The hybrid Wannier states in the opposite valley are related by time reversal symmetry. 
\subsection{Magnetic translation group eigenstates generated by hybrid Wannier states}
\begin{figure}
    \centering
    \includegraphics[width=0.9\linewidth]{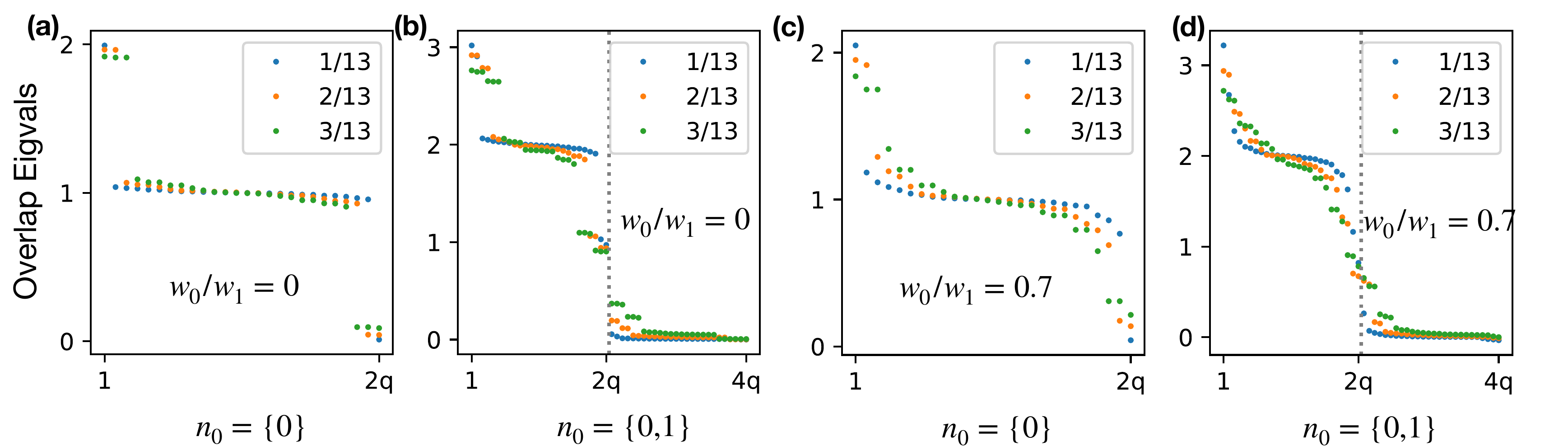}
    \caption{Eigenvalues of the overlap matrix $\Lambda_W(c_1l_1n_1,c_2l_2n_2;k_1,k_2)$ for a fixed $k_1=k_2=0$. Parameter choices are $q=13$ and $\theta=1.05^\circ$. (a,b) correspond to the chiral limit $w_0/w_1=0$, and (c,d) correspond to $w_0/w_1=0.7$.}
    \label{fig:figS2}
\end{figure}
In finite magnetic field, we choose the Landau gauge $\mathbf{A}=|\bB|x\hat{\by}$, and generate eigenstates of the magnetic translation group (MTG) via:
\begin{equation}\label{eq:mtg_basis}
    \ket{W_{c}(k_1,k_2;n_0)} =  \frac{1}{\sqrt{N}}\sum_{s_1=-\infty}^{\infty} e^{i2\pi k_1 s_1}\hat{t}_{\bL_1}^{s_1} \ket{w_c(n_0,k_2\bg_2)},
\end{equation}
where $\hat{t}_{\bL_{1,2}}$ are generators of magnetic translations by the primitive vectors of the moire unit cell $\bL_1=L_m(\frac{\sqrt{3}}{2},\frac{1}{2})$ and $\bL_2=L_m(0,1)$. They are given by:
\begin{equation}
    \hat{t}_{\bL_1} = e^{i\bq_\phi\cdot \br} \hat{T}_{\bL_1},\ \hat{t}_{\bL_2} = \hat{T}_{\bL_2},
\end{equation}
where $\hat{T}_{\bL_{1,2}}$ are usual discrete translation operators, defined via their action on a function $\psi(\br)$ as $\hat{T}_{\bL_{1,2}}\psi(\br)=\psi(\br-\bL_{1,2})$, and we have defined a magnetic translation wavevector:
\begin{equation}
    \bq_\phi = \frac{\phi}{\phi_0} \left(\frac{1}{2}\bg_1+\bg_2\right).
\end{equation}

For rational fluxes $\phi/\phi_0 = p/q$, it is straightforward to show that $\comm{\hat{t}_{\bL_1}}{\hat{t}_{\bL_2}^q}=0$, and:
\begin{equation}
    \hat{t}_{\bL_1}\ket{W_{c}(k_1,k_2;n_0)} = e^{-i2\pi k_1}\ket{W_{c}(k_1,k_2;n_0)},\  \hat{t}_{\bL_2}^q\ket{W_{c}(k_1,k_2;n_0)} = e^{-i2\pi q k_2}\ket{W_{c}(k_1,k_2;n_0)}.
\end{equation}
Note however that in general $\comm{\hat{t}_{\bL_1}}{\hat{t}_{\bL_2}}\neq 0$. We can relabel the MTG eigenstates as:
\begin{equation}
    \ket{W_c(k_1,k_2+\frac{l}{q};n_0)},\ k_1\in[0,1),\ k_2 \in[0,1/q),\ l=0,\dots q-1. 
\end{equation}
Here the magnetic Brillouin zone is defined by the magnetic strip $[0,1)\times[0,1/q)$. MTG eigenstates labeled by different $k_1$,$k_2$ quantum numbers in the magnetic Brillouin zone are orthogonal. For a given index $n_0$, there are $2q$ states generated from hybrid Wannier states, and are labeled by the Chern number $c$ and the index $l$ of magnetic strips along the $\bg_2$ direction.

Due to the nontrivial band topology encoded in the hybrid Wannier wavefunctions, the MTG eigenstates defined in Eq.~(\ref{eq:mtg_basis}) are not guaranteed to be orthonormal. Consider the overlap matrix at a given momentum $k_1$,$k_2$:
\begin{equation}\label{eq:M}
    \begin{split}
        & \Lambda_W(c_1l_1n_1,c_2l_2n_2;k_1,k_2) \\
        \equiv &  \bra{W_{c_1}(k_1,k_2+\frac{l_1}{q};n_1)}\ket{W_{c_2}(k_1,k_2+\frac{l_2}{q};n_2)}\\
        =  &\sum_{s_1} e^{i2\pi k_1 s_1} \bra{w_{c_1}(n_1,(k_2+\frac{l_1}{q})\bg_2)}\hat{t}_{\bL_1}^{s_1}\ket{w_{c_2}(n_2,(k_2+\frac{l_2}{q})\bg_2)}\\
        = & \sum_{s_1} e^{i2\pi k_1 s_1}e^{-i\frac{s_1(s_1-1)}{2}\bq_\phi\cdot \bL_1} \bra{w_{c_1}(n_1,(k_2+\frac{l_1}{q})\bg_2)}e^{is_1\bq_\phi\cdot \br}\ket{w_{c_2}(n_2+s_1,(k_2+\frac{l_2}{q})\bg_2)}\\
        =& \frac{1}{N_1}\sum_{s_1,\bar{k}_1,\bar{p}_1}e^{i2\pi k_1 s_1}e^{-i\frac{s_1(s_1-1)}{2}\bq_\phi\cdot \bL_1} e^{-i2\pi \bar{k}_1n_1}e^{i2\pi \bar{p}_1(n_2+s_1)} \bra{\Psi_{c_1}(\bar{k}_1\bg_1,(k_2+\frac{l_1}{q})\bg_2)}e^{is_1\bq_\phi\cdot \br}\ket{\Psi_{c_2}(\bar{p}_1\bg_1,(k_2+\frac{l_2}{q})\bg_2)}\\
        \equiv& \frac{1}{N_1}\sum_{s_1,\bar{k}_1,\bar{p}_1}e^{i2\pi k_1 s_1}e^{-i\frac{s_1(s_1-1)}{2}\bq_\phi\cdot \bL_1} e^{-i2\pi \bar{k}_1n_1}e^{i2\pi \bar{p}_1(n_2+s_1)} M(c_1\bar{\bk},c_2\bar{\bp};s_1),
    \end{split}
\end{equation}
where on the last line for notational simplicity we have defined $\bar{\bk}=\bar{k}_1\bg_1+(k_2+\frac{l_1}{q})\bg_2$, $\bar{\bp}=\bar{p}_1\bg_1+(k_2+\frac{l_2}{q})\bg_2$, and $M(c_1\bar{\bk},c_2\bar{\bp};s_1)$ as the matrix elements of $e^{i s_1 \bq_\phi\cdot \br}$ in the Chern Bloch basis. The magnetic wavevector $\bq_\phi$ hybridizes Chern Bloch states at different wavevectors satisfying the following two Diophantine equations:
\begin{equation}
    \bar{k}_1 = [\bar{p}_1 + s_1\frac{p}{2q}]_{1},\ l_1 = [l_2+s_1 p]_q,
\end{equation}
where $[\dots]_n$ denotes the modulus with respect to integer $n$.

\begin{figure}
    \centering
    \includegraphics[width=0.65\linewidth]{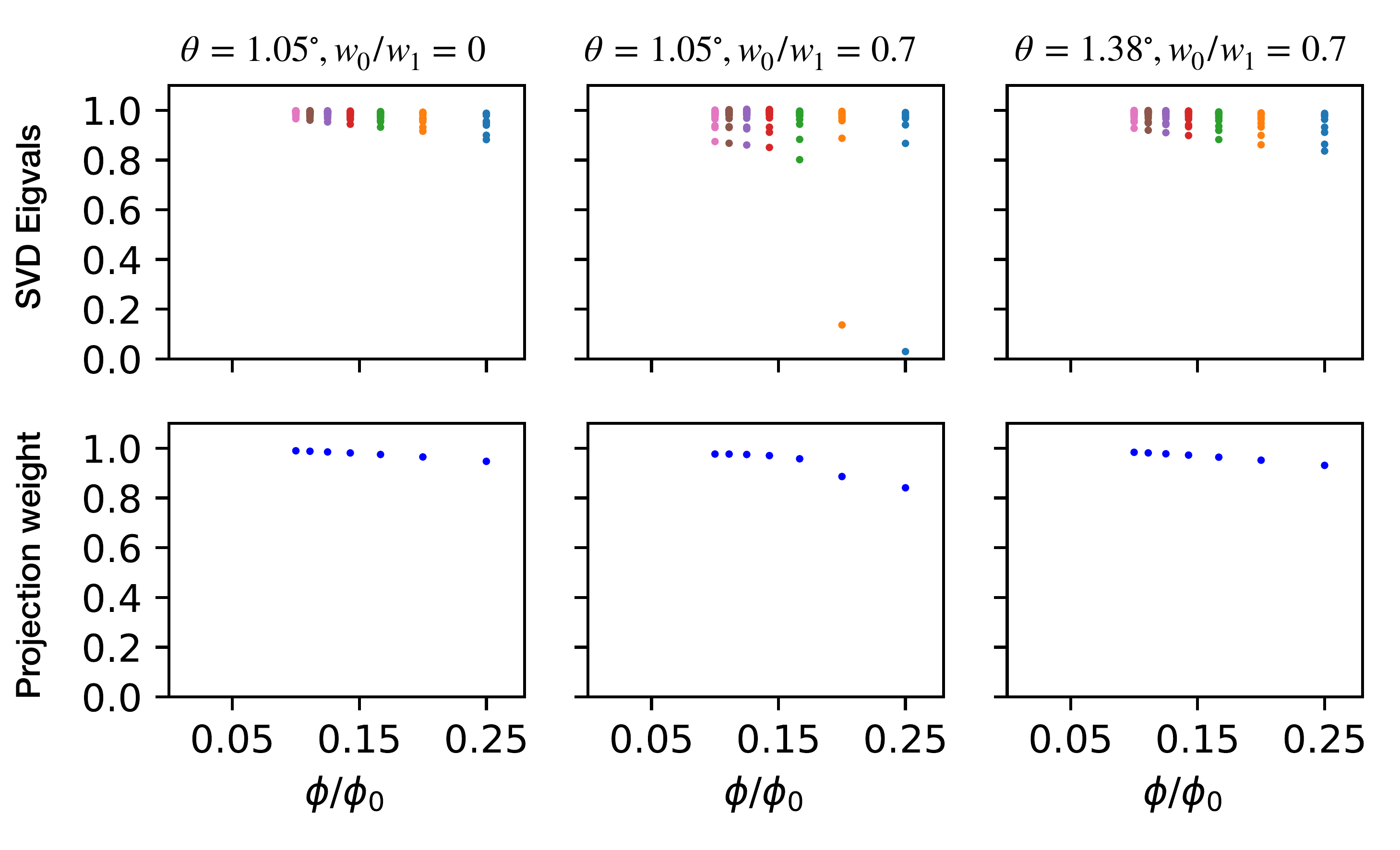}
    \caption{Measures of the goodness of hybrid Wannier approach on representing the exact $\mathbf{B}\neq 0 $ narrow band wavefunctions for a few values of $w_0/w_1$ , twist angle $\theta$ at $\bk=(0,0)$ of the magnetic Brillouin zone. Top panel, the SVD eigenvalues of the overlap matrix $\Lambda_{LL}(a,n;k_1,k_2)\equiv \bra{V_a(k_1,k_2)}\ket{\Psi_n(k_1,k_2)}$, where $\ket{\Psi_n(k_1,k_2)}$ ($n=1\dots 2q$) are exact wavefunctions calculated from the Landau-level wavefunction based approach. Bottom panel, the weight of $\ket{V_a(k_1,k_2)}$ inside the narrow bands, defined as $\text{Tr}[\Lambda_{LL}\Lambda_{LL}^\dagger]/2q$. Weight $=1$ indicates that $\ket{V_a(k_1,k_2)}$ is related to the exact narrow band wavefunctions via a unitary rotation. The exact $\bB\neq 0$ narrow band eigenstates  using the Landau level approach are calculated with upper cutoff on the Landau level index $n_{LL}=25q$. }
    \label{fig:figS3} 
\end{figure}

We first discuss the completeness of the basis states. Note that if we were to fix the indices $n_1=n_2=0$, the $2q$ MTG eigenstates are not linearly independent. This is illustrated in Fig.~\ref{fig:figS2}(a,c), where we show the eigenvalues of $\Lambda_W$ at momentum $k_1=k_2=0$. At $\phi/\phi_0=p/q$, there are $2q-p$ linearly independent states. An analytical proof of the state-deficiency is given in Sec. I of the SM for the chiral limit $w_0/w_1=0$ using an index theorem. On the other hand, a complete basis set can be generated by enlarging the $n_{1,2}$ range and choosing $n_1,n_2\in\{0,1\}$ for the trial MTG eigenstates. As illustrated in Fig.~\ref{fig:figS2}(b,d), this procedure generates an overcomplete basis set (i.e., number of independent states greater than $2q$). We therefore choose  $2q$ states with largest overlap eigenvalues to represent the narrow band Hilbert space in a finite magnetic field. Note however, that only at low magnetic fields are the $2q$ largest eigenvalues well separated from the remaining $2q$ states by a well defined spectral gap, i.e. the spillover from remote bands is small at low magnetic fields. However, at larger $\bB$ the overlap spectral gap closes (e.g. Fig.~\ref{fig:figS2}(d) with $\frac{\phi}{\phi_0}=\frac{2}{13}$), due to significant spillover from the remote bands, making the $\bB=0$ hybrid Wannier construction less reliable. This is also seen in Fig.~\ref{fig:figS4} where we make a quantitative comparison of the Hofstadter spectra for the non-interacting BM Hamiltonian calculated using the hybrid Wannier approach and more conventional Landau-level based approach.

The desired orthonormalized basis set $\{ \ket{V_a(k_1,k_2)},\ a=1,\dots 2q\}$ is thus obtained via:
\begin{equation}
    \ket{V_a(k_1,k_2)} \equiv \sum_{c,l,n_0}\ket{W_c(k_1,k_2+\frac{l}{q};n_0)} U(cln_0,a;k_1,k_2)\frac{1}{\sqrt{D_a(k_1,k_2)}},
\end{equation}
where $D_a$ are the largest $2q$ eigenvalues of $\Lambda_W$ for any given $k_1$,$k_2$, and $U$ is a rectangular matrix satisfying:
\begin{equation}
   \sum_{c_1l_1n_1,c_2l_2n_2} U^*(c_1l_1n_1,a;k_1,k_2)\Lambda_W(c_1l_1n_1,c_2l_2n_2;k_1,k_2)U(c_2l_2n_2,b;k_1,k_2) = \delta_{a,b} D_a(k_1,k_2).
\end{equation}
One can straightforwardly check the orthonormality condition $\bra{V_a(k_1,k_2)}\ket{V_b(k_1',k_2')}=\delta_{k_1,k_1'}\delta_{k_2,k_2'}\delta_{a,b}$.

To quantify how well $\ket{V_a(k_1,k_2)}$ describes the $\bB\neq 0$ narrow band Hilbert space, we follow Ref.~\cite{Hejazi2019} and obtain the narrow band eigenstates $\ket{\Psi_n(k_1,k_2)}$ by expanding the BM Hamiltonian in the Landau level basis of monolayer graphene. We define the overlap matrix between states generated via these two procedures : 
\begin{equation}
    \Lambda_{LL}(a,n;k_1,k_2) \equiv \bra{V_{a}(k_1,k_2)}\ket{\Psi_n(k_1,k_2)}.
\end{equation}
If the hybrid Wannier approach generates exact eigenstates, then for each $k_1$ and $k_2$ the matrix $\Lambda_{LL}(a,n;k_1,k_2)$ is unitary of size $2q\times 2q$. In Fig.~\ref{fig:figS3} we show both the SVD eigenvalues of $\Lambda_{LL}$ and the projected weight ($\text{Tr}[\Lambda_{LL}\Lambda_{LL}^\dagger]/2q$) for a few twist angles and ratios of $w_0/w_1$. Observe that as the magnetic field decreases, the spillover of $\ket{V_a}$ into remote bands also decreases, and $\ket{V_a}$ extrapolates to the exact narrow band wavefunctions in the $\bB\rightarrow 0$ limit.

\subsection{Matrix elements of the non-interacting BM Hamiltonian in the MTG eigenstates}

The matrix elements of the BM Hamiltonian in the orthonormalized MTG eigenstate basis is given by:
\begin{equation}
\begin{split}
    & \bra{V_a(k_1,k_2)}\hat{H}^{\bK}_{BM}(\bp -\frac{e}{c}\bA) \ket{V_b(k_1,k_2)}\\
    = & \frac{1}{\sqrt{D_a(k_1,k_2)}}U^\dagger(a,c_1l_1n_1;k_1,k_2) H_{BM}(c_1l_1n_1,c_2l_2n_2;k_1,k_2)U(c_2l_2n_2,b;k_1,k_2)\frac{1}{\sqrt{D_b(k_1,k_2)}},
\end{split}
\end{equation}
where repeated indices are summed over, and: 
\begin{equation}\label{eq:HBM}
\begin{split} 
    & H_{BM}(c_1l_1n_1,c_2l_2n_2;k_1,k_2) \\
    \equiv & \bra{W_{c_1}(k_1,k_2+\frac{l_1}{q};n_1)} \hat{H}^{\bK}_{BM}(\bp -\frac{e}{c}\bA)\ket{W_{c_2}(k_1,k_2+\frac{l_2}{q};n_2)}\\
    = & \sum_{s_1} e^{i2\pi k_1 s_1}e^{-i\frac{s_1(s_1-1)}{2}\bq_\phi\cdot \bL_1} \bra{w_{c_1}(n_1,(k_2+\frac{l_1}{q})\bg_2)}\hat{H}^{\bK}_{BM}(\bp -\frac{e}{c}\bA)e^{is_1\bq_\phi\cdot \br}\ket{w_{c_2}(n_2+s_1,(k_2+\frac{l_2}{q})\bg_2)}.
\end{split}
\end{equation}

The matrix elements of the BM Hamiltonian can be split into two terms
\begin{equation} \label{eq:HBMmatelement}
\hat{H}^{\bK}_{BM}(\bp-\frac{e}{c}\bA) = \hat{H}^{\bK}_{BM}(\bp)- \frac{ \hbar v_F}{\ell}(1_2\sigma_y)\frac{x}{\ell},
\end{equation}
where $\ell\equiv \sqrt{\frac{\hbar c}{e B}}$ is the magnetic length. 

We stress that due to exponential localization of the hybrid Wannier states along the $\bL_1$ direction, the matrix elements of the BM Hamiltonian in Eq.~(\ref{eq:HBM}) are non-neglibile only if $n_1$ and $n_2+s_1$ are close to each other. Since both $n_1,n_2\in\{0,1\}$, this constrains the summation over $s_1$ to a few moir\'e lattice constants. In practice we are able to achieve numerical convergence for $s_1\in [-4,4]$.  Therefore, $x$ in the matrix element is $O(L_m)$ making the second term $O(\hbar v_F L_m/\ell^2)$. The first term in Eq.~(\ref{eq:HBMmatelement}) is nominally $O(\hbar v_F k_\theta)$ except very near the magic angle where there is an additional suppression of the bandwidth by a factor of $\eta\sim 1/40$ (in the chiral limit magic angle $\eta$ vanishes). The second term is therefore nominally smaller than the first term at weak magnetic fields by a factor of order $O(L^2_m/(4\ell^2))$, except near the magic angle where this factor has an extra enhancement by  $1/\eta$. 

The matrix elements of the $\bB=0$ term are calculated as follows:
\begin{equation}
\begin{split}
    & H_{BM}^{(1)}(c_1l_1n_1,c_2l_2n_2;k_1,k_2) \\
    = & \frac{1}{N_1}\sum_{s_1 \bar{k}_1\bar{p}_1}e^{i2\pi k_1 s_1}e^{-i\frac{s_1(s_1-1)}{2}\bq_\phi\cdot \bL_1} e^{-i2\pi \bar{k}_1n_1}e^{i2\pi \bar{p}_1(n_2+s_1)}\sum_{c_3}\varepsilon_{c_1,c_3}(\bar{\bk})M(c_1\bar{\bk},c_2\bar{\bp};s_1),
\end{split}
\end{equation}
where $\varepsilon_{c_1,c_3}(\bar{\bk})$ is the matrix elements of the zero field BM Hamiltonian in the Chern Bloch basis, and $M$ is defined in Eq.~(\ref{eq:M}).

The matrix elements for the vector potential term is calculated as:
\begin{equation}
    \begin{split}
        & H_{BM}^{(2)}(c_1l_1n_1,c_2l_2n_2;k_1,k_2)\\
    = & -\frac{\hbar v_F}{\ell}\frac{1}{N_1}\sum_{s_1\bar{k}_1\bar{p}_1} e^{i2\pi k_1 s_1}e^{-i\frac{s_1(s_1-1)}{2}\bq_\phi\cdot \bL_1} e^{-i2\pi \bar{k}_1n_1}e^{i2\pi \bar{p}_1(n_2+s_1)} N(c_1\bar{\bk},c_2\bar{\bp};s_1),
    \end{split}
\end{equation}
where we have defined:
\begin{equation} \label{eq:N}
    N(c_1\bar{\bk},c_2\bar{\bp};s_1) \equiv \frac{1}{N_1L_{1,x}}\frac{1}{qN_2L_{2}}\int_{-N_1L_{1,x}/2}^{N_1L_{1,x}/2}\mathrm{d}x \int_{0}^{qN_2L_{2}}\mathrm{d}y  \Psi_{c_1\bar{\bk}}^\dagger(\br)\left[\sigma_y\mu_0 \frac{x}{\ell}e^{is_1\bq_\phi\cdot \br} \right]\Psi_{c_2\bar{\bp}}(\br).
\end{equation}
Here we write it explicitly as a real space integral. We chose the momentum space mesh to be $k_1=\frac{m_1}{N_1}$ where $m_1=0,\dots,N_1-1$, and $k_2=\frac{m_2}{N_2q}$ where $m_2=0,\dots,N_2-1$. The $x$-integration range in Eq.~(\ref{eq:N}) is necessary to avoid revivals of the hybrid Wannier states since $\ket{w_{c}(n+N_1,k_2\bg_2)}=\ket{w_{c}(n,k_2\bg_2)}$. Eq.~(\ref{eq:N}) is calculated by expressing the Chern Bloch states in the plane wave basis, and perform real-space integration accordingly. 

\begin{figure}
    \centering
    \includegraphics[width=0.8\linewidth]{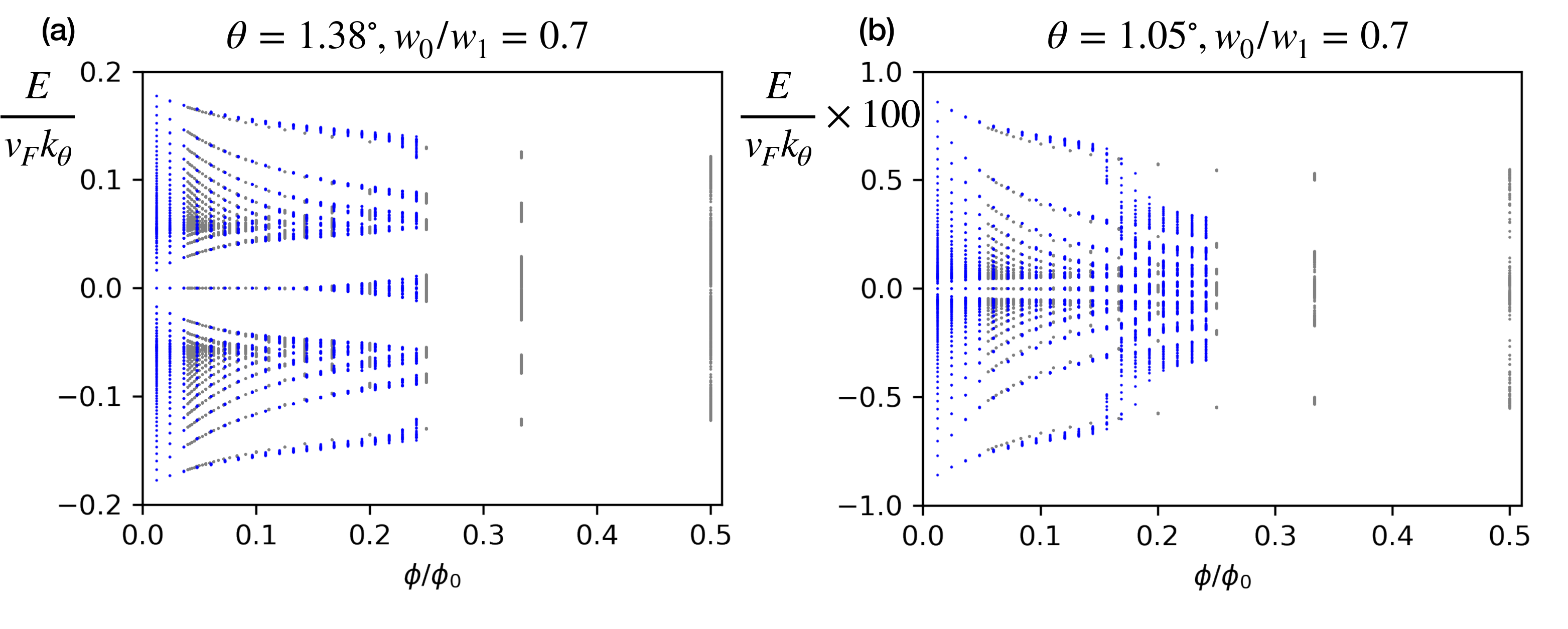}
    \caption{Comparison of the Hofstadter spectra of the non-interacting BM Hamiltonian calculated via the hybrid Wannier approach (blue) and conventional Landau level approach (gray). (a) is away from magic angle at $\theta=1.38^\circ$, and (b) is at the magic angle.}
    \label{fig:figS4}
\end{figure}

In Fig.~\ref{fig:figS4} we show the Hofstadter spectra calculated using the hybrid Wannier approach in comparison with the exact spectra calculated using the Landau level approach discussed in Ref.~\cite{Hejazi2019}. The spectra based on hybrid Wannier approach are calculated for $\phi/\phi_0=p/83$, where $p=1\dots 20$. For the Landau level approach, we only compute the $1/q$ sequence for $q=2\dots 18$. To achieve numerical convergence we choose the upper cutoff for Landau levels at $n_{LL}=25q$ (total number of Landau levels kept is $25q$ including the zeroth Landau level). Observe that the Hofstadter spectra calculated using the hybrid Wannier approach faithfully reproduce the exact results, with better agreement as the magnetic field is reduced. At larger magnetic fields, the spillover effect from the remote bands is significant, making the hybrid Wannier approach unreliable. This occurs at $\frac{\phi}{\phi_0}\approx 0.157$ for parameters used in Fig.~\ref{fig:figS4}(b), which coincides roughly where the spectral gap of the overlap matrix closes, shown in Fig.~\ref{fig:figS2}(d). Therefore, the spectral gap of the overlap matrix provides a measure of how good the $\bB=0$ hybrid Wannier approach is in describing the narrow band Hilbert space at $\bB\neq 0$.

We note that due to the exponential localization of the hybrid Wannier states along the $\bL_1$-direction, in practice when calculating the matrix elements in the hybrid Wannier approach, we restrict the summation over $s_1$ from $-4$ to $4$. We checked that numerical convergence has been achieved for $\Lambda_W$, matrix elements of BM Hamiltonian, as well as matrix elements of the strong coupling Hamiltonian. Furthermore we choose a momentum mesh such that $N_1=q N_2=2q n$ where we adjust integer $n$ to achieve numerical convergence with the momentum mesh.

\subsection{Matrix elements of the projected density operator}
In momentum space, the Coulomb interaction takes the form: 
\begin{equation}
    \hat{H}_{int} = \frac{1}{2}\sum_{\bq}V(\bq) \delta \rho_{\bq}\delta \rho_{-\bq}
\end{equation}
where the projected electron density operator in valley $\bK$ is given by:
\begin{equation}\label{eq::rhoq}
    \hat{\rho}_\bq = \sum_{a,k_1,k_2;b,p_1,p_2}\bra{V_a(k_1,k_2)}e^{i\bq \cdot \br} \ket{V_b(p_1,p_2)} \frak{d}^\dagger_{a,k_1,k_2}\frak{d}_{b,p_1,p_2}
\end{equation}
The density operator in valley $\bK'$ is related via $C_2P$ symmetry. 

The background charge term $\bar{\rho}_\bq$ can be calculated from valley $\bK$ (guaranteed by $C_2P$ symmetry), and is given as:
\begin{equation}\label{eq::rhoqbar}
    \bar{\rho}_\bq = 2\sum_{m,n\in \mathbb{Z}} \delta_{\bq,m\bg_1+n\bg_2} \sum_{a,k_1,k_2}\bra{V_a(k_1,k_2)}e^{i\bq\cdot \br}\ket{V_a(k_1,k_2)}
\end{equation}
where $2$ comes from spin degeneracy (neglecting Zeeman splitting). Eqs.~(\ref{eq::rhoq}) and (\ref{eq::rhoqbar}) are calculated in a similar fashion to the matrix elements of the BM Hamiltonian discussed previously. We choose $m,n\in \{ 0,\pm 1,\pm2,\pm3\}$ for numerical convergence.

\section{Solving for the dispersion of charged excitations in the strong coupling limit using Landau-level based approach}
\subsection{MTG eigenstates generated from Landau level states}
In the main text we have focused our attention on the hybrid Wannier approach and solving for the $\bB\neq 0$ dispersion of charged excitations. Here we also present a calculation based on the Landau-level approach~\cite{Bistritzer2011b,Moon2014,Hejazi2019,Yahui2019}, which is much more computationally demanding at low $\bB$. Later we provide a consistency check between Landau-level approach and the hybrid Wannier approach. 

We begin with a brief discussion of the Landau level eigenstates of the Dirac Hamiltonian of monolayer graphene. For simplicity we consider the following Dirac Hamiltonian in a magnetic field:
\begin{equation}
    \hat{H}^{\bK}_{l}(\bp - \frac{e\bA}{c}) =  v_F \left[\sigma_x (p_x-K_{l,x})+\sigma_y(p_y-K_{l,y}-x/\ell^2)\right].
\end{equation}
Here $l=1,2$ is the layer index, and $\bK_l=(K_{l,x},K_{l,y})$ is the position of the Dirac cone in the reciprocal space. The eigenstates of the Dirac Hamiltonian are solved by going to the harmonic oscillator basis: $x=\frac{\ell}{\sqrt{2}}(a+a^\dagger)$, and $p_x=\frac{1}{i\sqrt{2}\ell}(a-a^\dagger)$. The particle-hole symmetric Landau level eigenstates are given as:
\begin{equation}
    \bra{\br} \ket{\psi^{(l)}_{n\gamma}(k_2)} = e^{iK_{l,x}x}e^{i\frac{2\pi}{L_m} k_2y} \frac{1}{\sqrt{2}} \begin{pmatrix}\phi_{n}(x-\tilde{k}_{2,l}\ell^2)\\ -i\gamma \phi_{n-1}(x-\tilde{k}_{2,l}\ell^2)\end{pmatrix},
\end{equation}
where $\epsilon_{n\gamma}=\frac{v_F}{\ell}\gamma \sqrt{2n}$ is the energy of the Dirac Hamiltonian, labeled by $n=1,2,\dots$, and $\gamma=\pm 1$ corresponds to positive and negative energy solutions. In addition, there is an anomalous zero energy state given by
\begin{equation}
   \bra{\br} \ket{\psi^{(l)}_{0}(k_2)} = e^{iK_{l,x}x}e^{i\frac{2\pi}{L_m} k_2y}  \begin{pmatrix}\phi_{n}(x-\tilde{k}_{2,l}\ell^2)\\ 0\end{pmatrix}
\end{equation}
which lives on the A sublattice. $\phi_n(x)$ is the eigenfunction of $a^\dagger a$, and is given to be:
\begin{equation}
    \phi_n(x) = \frac{1}{\pi^{1/4}}\frac{1}{\sqrt{2^nn!}}e^{-x^2/2\ell^2}H_n(x/\ell),
\end{equation}
where $H_n(x)$ is the Hermite polynomial. The shift in the position for a given momentum $k_2\bg_2$ is given by:
\begin{equation}
    \tilde{k}_{2,l}\ell^2 = \left(\frac{2\pi}{L_m}k_2 - K_{l,y}\right)\ell^2.
\end{equation}

In the Landau level basis, the eigenstates of the MTG are generated as:
\begin{equation}
\begin{split}
    \ket{\Psi_{n\gamma}^{(l)}(k_1,k_2)} & = \sum_{s_1=-\infty}^{\infty}e^{i2\pi k_1 s_1} \hat{t}_{\bL_1}^{s_1} \ket{\psi^{(l)}_{n\gamma}(k_2)} \\
    & = \sum_{s_1}e^{i2\pi (k_1-\frac{k_2}{2})s_1}e^{-i\frac{s_1(s_1-1)}{2}\bq_\phi\cdot \bL_1}e^{-is_1K_{l,x}L_{1,x}} \ket{\psi^{(l)}_{n\gamma}(k_2+s_1\frac{\phi}{\phi_0})}.
\end{split}
\end{equation}
It is straightforward to check that:
\begin{align}
    \ket{\Psi_{n\gamma}^{(l)}(k_1+1,k_2)} & = \ket{\Psi_{n\gamma}^{(l)}(k_1,k_2)},\\
    \ket{\Psi_{n\gamma}^{(l)}(k_1,k_2+\frac{\phi}{\phi_0})} & = e^{iK_{l,x}L_{1,x}}e^{-i2\pi(k_1-\frac{k_2}{2})} \ket{\Psi_{n\gamma}^{(l)}(k_1,k_2)},\\
    \bra{\Psi_{n\gamma}^{(l)}(k_1,k_2)}\ket{\Psi_{n'\gamma'}^{(l')}(p_1,p_2)} & = \delta_{l,l'}\delta_{k_1,k_2}\delta_{p_1,p_2}\delta_{n,n'}\delta_{\gamma,\gamma'}.
\end{align}
Therefore, the MTG eigenstates defined in $(k_1,k_2)\in [0,1)\times[0,\frac{\phi}{\phi_0})$ form a complete and orthornomal basis set in a finite magnetic field. 

The exact eigenstates of the narrow bands for non-interacting BM Hamiltonian are solved by computing the matrix elements in these MTG eigenstates.

\subsection{Strong coupling Hofstadter spectra using Landau level approach}
We project the strong coupling Hamiltonian onto the $\bB\neq 0$ narrow band eigenstates, which we label as $\ket{\Psi_{a,\bk}}$ where $a=1,\dots 2q$, and $\bk=k_1\bg_1+k_2\bg_2$. The Hamiltonian can be written as:
\begin{align}
    \hat{H}_{int} & = \frac{1}{2} \sum V_\bq \delta\rho_\bq \delta\rho_{-\bq}, \\
    \delta \rho_\bq & = \sum_{\eta=\bK,\bK'}\sum_{s=\uparrow,\downarrow}\sum_{a\bk,b\bp} \bra{\Psi_{\eta s, a,\bk}}1_4e^{i\bq \cdot \br}\ket{\Psi_{\eta s, b,\bp}} \frak{d}^{\dagger}_{\eta s, a,\bk}\frak{d}_{\eta s, b,\bp} - \bar{\rho}_\bq,\\
    \bar{\rho}_\bq & = \frac{1}{2} \sum_{m,n\in \mathbb{Z}} \delta_{\bq,m\bg_1+n\bg_2}\sum_{\eta,s}\sum_{a,\bk}  \bra{\Psi_{\eta s, a,\bk}}1_4e^{i\bq \cdot \br}\ket{\Psi_{\eta s, a,\bk}}.\label{eq:rhoq}
\end{align}
Here $1_4$ is the identity operator in the Hilbert space spanned by layer and sublattice degrees of freedom. For completeness we added the valley ($\eta$) and spin ($s$) indices as subscripts to the narrow band eigenstates. The background charge density term $\bar{\rho}_\bq$ can be calculated entirely in valley $\bK$, due to the $C_2P$ symmetry relating the eigenstate wavefunctions in the two valleys. To obtain the energy spectrum of excitations at integer fillings we apply the double commutator method discussed in the main text. 

The key numerical procedure is to calculate the overlap matrix $\bra{\Psi_{a,\bk}}1_4 e^{i\bq \cdot \br}\ket{\Psi_{b,\bp}}$, which we discuss below. From here onward we also drop the valley and spin indices for notational convenience. We proceed by constructing the matrix in the MTG eigenstate basis:
\begin{equation} \label{eq:LLstrongcoupling}
    \bra{\Psi^{(l_1)}_{n_1\gamma_1,\bk}} 1_4 e^{i\bq \cdot \br}\ket{\Psi^{(l_2)}_{n_2\gamma_2,\bp}},
\end{equation}
and then project onto the narrow bands via:
\begin{equation} 
   \bra{\Psi_{a,\bk}}1_4e^{i\bq \cdot \br}\ket{\Psi_{b,\bp}} =   U^\dagger_{a,n_1\gamma_1l_1}(\bk)\bra{\Psi^{(l_1)}_{n_1\gamma_1,\bk}}1_4 e^{i\bq \cdot \br}\ket{\Psi^{(l_2)}_{n_2\gamma_2,\bp}}U_{n_2\gamma_2l_2,b}(\bp),
\end{equation}
where $U$ is a rectangular part of the unitary matrix that diagonalizes the BM Hamiltonian. Repeated indices are summed over. 

The matrix element defined in Eq.~(\ref{eq:LLstrongcoupling}) is calculated as follows:
\begin{equation}
    \begin{split}
        & \bra{\Psi^{(l_1)}_{n_1\gamma_1,\bk}} 1_4 e^{i\bq \cdot \br}\ket{\Psi^{(l_2)}_{n_2\gamma_2,\bp}}\\
        = &\sum_{s_1,s_1'}e^{-i2\pi k_1 s_1}e^{i2\pi p_1 s_1'}e^{i2\pi q_1s_1}\bra{\psi^{(l_1)}_{n_1\gamma_1}(k_2)} 1_4 e^{i\bq \cdot \br}\hat{t}_{\bL_1}^{s_1'-s_1}\ket{\psi^{(l_2)}_{n_2\gamma_2}(p_2)}\\
        =& \sum_{s_1}\delta_{[p_1+q_1]_1,k_1}e^{i2\pi p_1 s_1} \bra{\psi^{(l_1)}_{n_1\gamma_1}(k_2)} 1_4 e^{i\bq \cdot \br}\hat{t}_{\bL_1}^{s_1}\ket{\psi^{(l_2)}_{n_2\gamma_2}(p_2)}\\
        = &\delta_{l_1,l_2}\sum_{s_1}\delta_{[p_1+q_1]_1,k_1}e^{i2\pi (p_1-\frac{p_2}{2}) s_1} e^{-i\frac{s_1(s_1-1)}{2}\bq_\phi\cdot \bL_1} e^{-is_1K_{l,x}L_{1,x}}\bra{\psi^{(l_1)}_{n_1\gamma_1}(k_2)} 1_2 e^{i\bq \cdot \br}\ket{\psi^{(l_1)}_{n_2\gamma_2}(p_2+s_1\frac{\phi}{\phi_0})}.
    \end{split}
\end{equation} 

Here we only calculate the expression when both Landau level indices $n_1$ and $n_2$ are non-zero. The case where either is zero can be calculated straightforwardly. The expectation value of $1_2e^{i\bq\cdot \br}$ operator in the Landau level basis of layer $l$ is given as follows:
\begin{equation}
    \begin{split}
        &  \bra{\psi^{(l)}_{n_1\gamma_1}(k_2)} 1_2 e^{i\bq \cdot \br}\ket{\psi^{(l)}_{n_2\gamma_2}(p_2+s_1\phi/\phi_0)}\\
   = & \frac{1}{2}\delta_{k_2,p_2+s_1\frac{\phi}{\phi_0}+q_2}\int_{-\infty}^{\infty}\mathrm{d}x \begin{pmatrix}\phi_{n_1}(x-\tilde{k}_{2,l}\ell^2), & i\gamma_1 \phi_{n_1-1}(x-\tilde{k}_{2,l}\ell^2)\end{pmatrix} 1_2 e^{iq_x x} \begin{pmatrix} \phi_{n_2}(x-(\tilde{k}_{2,l}-q_y)\ell^2) \\ -i\gamma_2 \phi_{n_2-1}(x-(\tilde{k}_{2,l}-q_y)\ell^2)\end{pmatrix} \\
   = & \frac{1}{2}\delta_{k_2,p_2+s_1\frac{\phi}{\phi_0}+q_2} \left[\int_{-\infty}^{\infty}\mathrm{d}x \phi_{n_1}(x-\tilde{k}_{2,l}\ell^2)e^{iq_x x} \phi_{n_2}(x-(\tilde{k}_{2,l}-q_y)\ell^2)+ \right.\\
   & \left. \gamma_1\gamma_2\int_{-\infty}^{\infty}\mathrm{d}x  \phi_{n_1-1}(x-\tilde{k}_{2,l}\ell^2)e^{iq_x x} \phi_{n_2-1}(x-(\tilde{k}_{2,l}-q_y)\ell^2)\right].
    \end{split}
\end{equation}

Note that:
\begin{equation}
    \begin{split}
        & \int_{-\infty}^{\infty}\mathrm{d}x\phi_{n_1}(x-\tilde{k}_{2,l}\ell^2)e^{iq_x x} \phi_{n_2}(x-(\tilde{k}_{2,l}-q_y)\ell^2)\\
    = & e^{iq_x \tilde{k}_{2,l}\ell^2}\int_{-\infty}^{\infty}\mathrm{d}x {\phi_{n_1}(x)}e^{iq_x x} {\phi_{n_2}(x+q_y\ell^2)}\\
    = & e^{iq_x \tilde{k}_{2,l}\ell^2}\int_{-\infty}^{\infty}\mathrm{d}x \phi_{n_1}(x)e^{iq_x x}e^{ip_x q_y\ell^2} \phi_{n_2}(x)\\
    = & e^{iq_x \tilde{k}_{2,l}\ell^2}e^{-\frac{i}{2}q_xq_y\ell^2}\int_{-\infty}^{\infty}\mathrm{d}x \phi_{n_1}(x)e^{iq_x x+iq_y\ell^2 p_x }\phi_{n_2}(x)\\
    = & e^{iq_x \tilde{k}_{2,l}\ell^2}e^{-\frac{i}{2}q_xq_y\ell^2}\int_{-\infty}^{\infty}\mathrm{d}x \phi_{n_1}(x)e^{c_- a+c_+a^\dagger} \phi_{n_2}(x),
    \end{split}
\end{equation}
where we have used $e^{X}e^{Y}=e^{X+Y+\frac{1}{2}\comm{X}{Y}+\dots}$, and:
\begin{equation}
    c_{\pm} = i\frac{\ell}{\sqrt{2}}(q_x\pm i q_y).
\end{equation}

The expectation value in the harmonic oscillator basis is calculated as follows:
\begin{equation}
\int_{-\infty}^{\infty}\mathrm{d}x {\phi_{n}(x)}e^{c_- a+c_+a^\dagger}{\phi_{m}(x)} = 
\begin{cases}
    e^{\frac{1}{2}c_+c_-}\sqrt{\frac{m!}{n!}}(c_+)^{n-m}L_m^{n-m}(-c_+c_-) &  \text{ for } n\ge m,\\
    e^{\frac{1}{2}c_+c_-}\sqrt{\frac{n!}{m!}}(c_-)^{m-n}L_n^{m-n}(-c_+c_-) &  \text{ for } n< m,
\end{cases}
\end{equation}
where $L_{n}^{m-n}(x)$ is the associated Laguerre polynomial,
\begin{eqnarray}
L^m_N(x)&=&\sum_{k=0}^N\frac{(N+m)!}{(N-k)!(m+k)!k!}(-x)^k.
\end{eqnarray}

\begin{figure}
    \centering
    \includegraphics[width=0.6\linewidth]{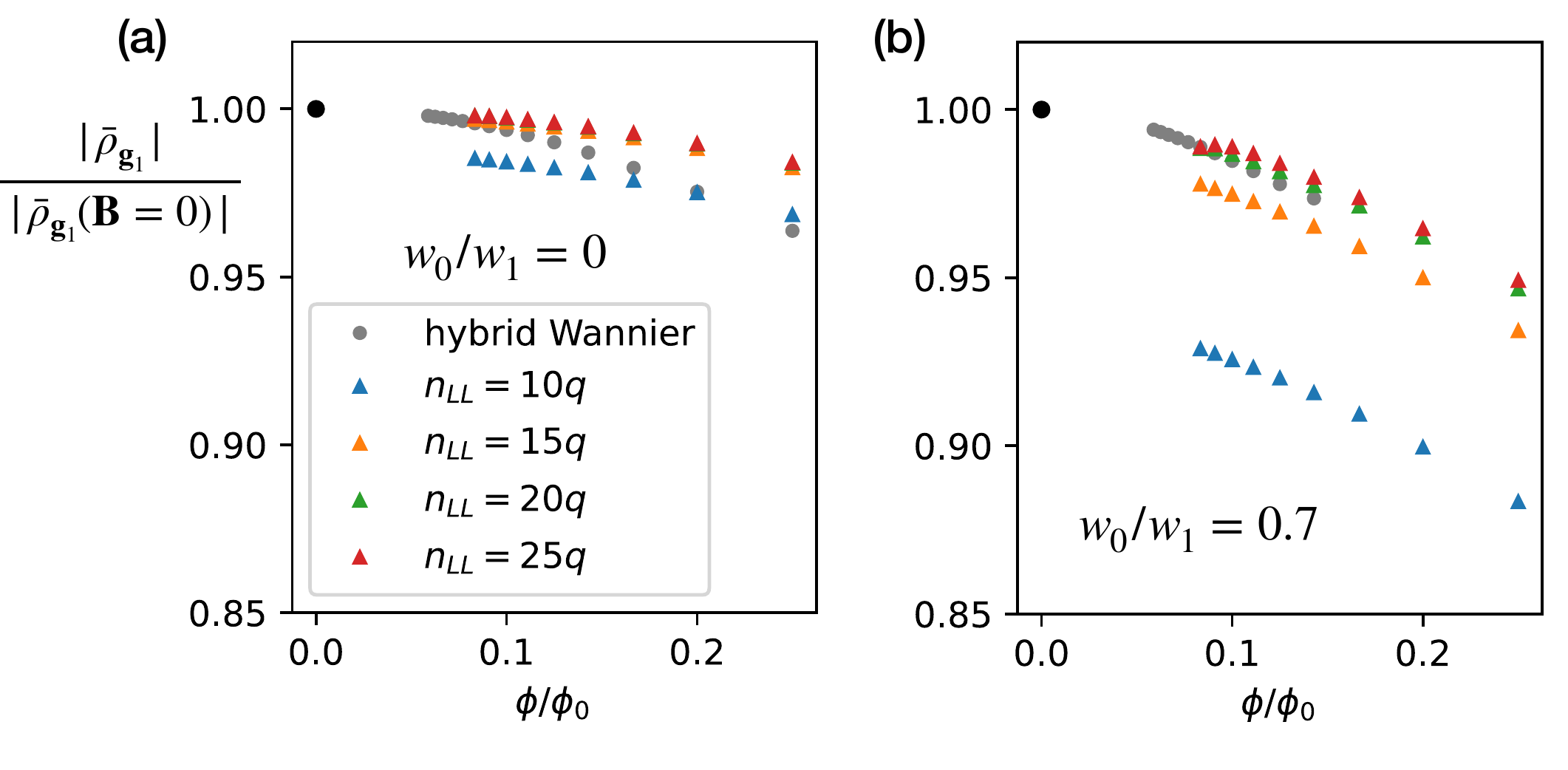}
    \caption{$|\bar{\rho}_{\bg_1}|$ normalized by its $\bB=0$ value, calculated using hybrid Wannier approach (gray) and Landau level approach with various upper Landau level cutoffs. }
    \label{fig:figS5}
\end{figure}
\subsection{Comparing Landau level  approach and hybrid Wannier approach}
As has been pointed out by Ref.~\cite{Hejazi2019}, to achieve numerical convergence for the Hofstadter spectra of the non-interacting BM Hamiltonian, the number of Landau levels to be kept is roughly $25q$, when $\phi/\phi_0=1/q$. This makes the calculation of strong coupling Hofstadter spectra prohibitively costly at low magnetic fields. On the other hand, as has already been demonstrated earlier in Fig.~\ref{fig:figS4}, in the hybrid Wannier approach, only two sets of Chern states are needed, making it much more computationally viable. 

In Fig.~\ref{fig:figS5} we show the magnetic field dependence of $|\bar{\rho}_{\bg_1}|$ defined in Eq.~(\ref{eq:rhoq}), calculated using the Landau-level approach for a sequence of upper Landau level cutoffs $n_{LL}$, as well as the hybrid Wannier approach. This shows that to achieve numerical convergence in the Landau level approach a significant number of Landau levels needs to be kept. Therefore, it is preferable to use the hybrid Wannier approach at low magnetic fields.

\begin{figure}
    \centering
    \includegraphics[width=0.9\linewidth]{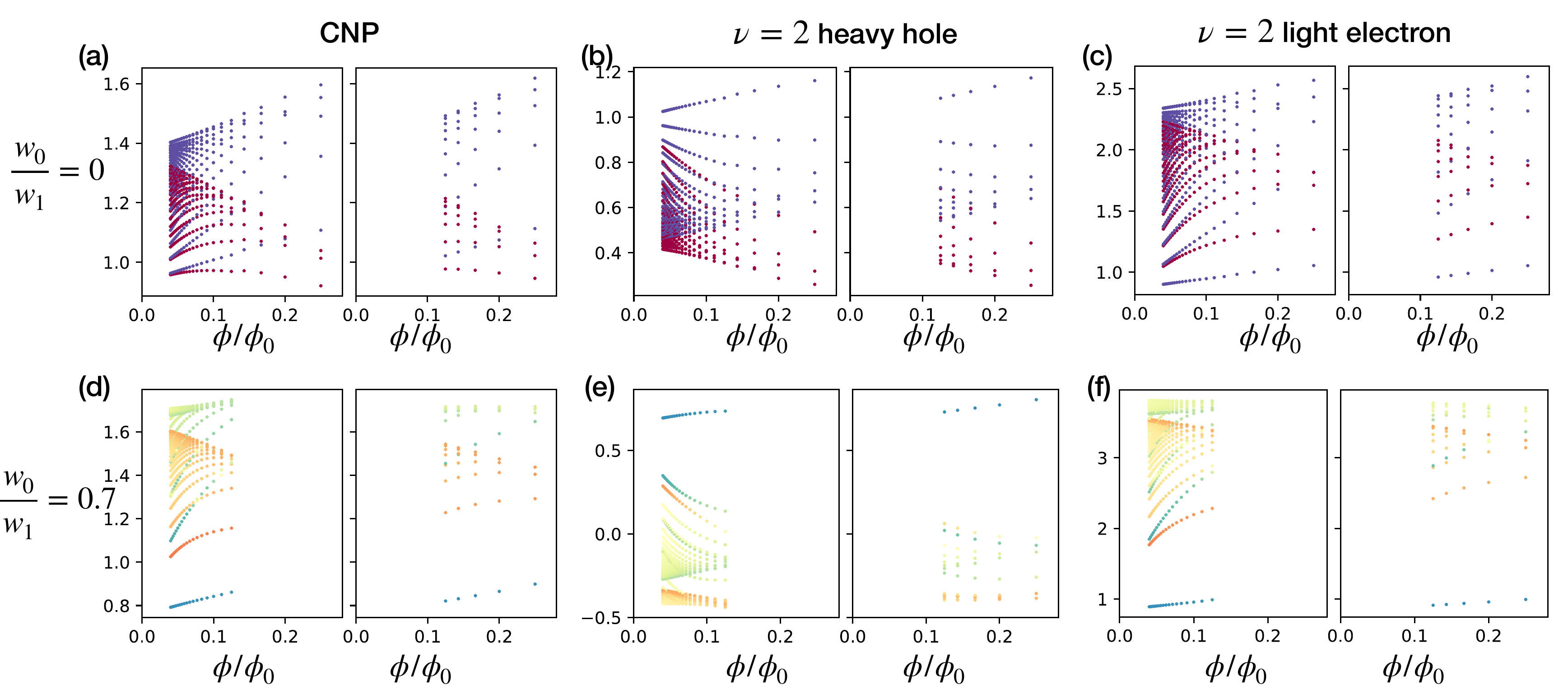}
    \caption{Comparing the strong coupling Hofstadter spectra at magic angle with  $w_0/w_1=0$ (a-c) and $w_0/w_1=0.7$ (d-f). In every figure, the left panel is the hybrid Wannier approach and right panel is Landau-level based approach. The three columns are charge $\pm 1$ excitations at charge neutral point, charge $+1$ at $\nu=2$, and charge $-1$ at $\nu=2$ respectively.}
    \label{fig:figS6}
\end{figure}

 In Fig.~\ref{fig:figS6} we compare the strong coupling Hofstadter spectra for these two approaches at a single momentum point $(k_1,k_2)=(0,0)$. The hybrid Wannier approach shows results for ${\phi}/{\phi_0}=1/4,\dots,1/25$. Due to numerical stability and convergence issues at low fields, we only show the Landau level approach calculation for ${\phi}/{\phi_0}=1/4,\dots,1/8$ with upper Landau level cutoff of $25q$. Due to remote band spillover effects at these values of $\phi/\phi_0$ for the hybrid Wannier approach (see for instance Fig.~\ref{fig:figS3}), full quantitative comparison should not be expected. Nevertheless, the two approaches display the same qualitative Landau quantization of the strong coupling energy dispersions, including Landau level degeneracies, the sublattice polarization, the opposite energetic evolution of the sublattice polarized bands with magnetic field close to the van Hove singularities of the $\bB=0$ dispersions, thereby providing an additional confirmation of the validity of the hybrid Wannier method introduced in this work.  

\end{document}